\documentclass[a4paper,11pt]{article}
\usepackage{jcappub} 
\usepackage[T1]{fontenc} 
\usepackage{graphics}
\title{\boldmath Entangled Scalar and Tensor Fluctuations during Inflation}
\author{Hael Collins}
\author{and Tereza Vardanyan}
\affiliation{Department of Physics, Carnegie Mellon University\\ 
5000 Forbes Avenue, Pittsburgh, Pennsylvania, U.S.A.}
\emailAdd{hcollins@andrew.cmu.edu}
\emailAdd{tvardany@andrew.cmu.edu}

\abstract{We show how the choice of an inflationary state that entangles scalar and tensor fluctuations affects the angular two-point correlation functions of the $T$, $E$, and $B$ modes of the cosmic microwave background.  The propagators for a state starting with some general quadratic entanglement are solved exactly, leading to predictions for the primordial scalar-scalar, tensor-tensor, and scalar-tensor power spectra.  These power spectra are expressed in terms of general functions that describe the entangling structure of the initial state relative to the standard Bunch-Davies vacuum.  We illustrate how such a state would modify the angular correlations in the CMB with a simple example where the initial state is a small perturbation away from the Bunch-Davies state.  Because the state breaks some of the rotational symmetries, the angular power spectra no longer need be strictly diagonal.}

\keywords{inflation, physics of the early universe, cosmological parameters from CMBR, cosmological perturbation theory}

\arxivnumber{1601.05415}

\begin{document}
\maketitle
\flushbottom
\setcounter{page}{2}

\section{Introduction} 

\noindent 
Experimental measurements of the early universe have become remarkably accurate and have allowed us to deduce the details of even earlier epochs \cite{Ade:2015lrj,Ade:2015ava}.  During the past few decades, a rich host of inflationary models have been proposed to explain the origin of this primordial structure.  While these models all must necessarily generate the same basic patterns in the primordial fluctuations, they can differ tellingly in what they predict for the yet finer structures in the fluctuations.  Different models can generate distinctive features in the primordial power spectra such as small oscillations about a largely scale invariant form, and they can vary quite substantially in the sizes and shapes of the elusive non-Gaussianities that they produce.  Choosing to preserve or to break the symmetries of the inflationary background determines further whether the resulting fluctuations are statistically isotropic or whether they can reproduce some of the oddities and anomalies seen in the low multipole moments of the microwave background radiation.

In contrast to the wealth of inflationary models, which are implemented through a choice of an action for the inflaton and perhaps other fields, much less work has been done to study the other essential ingredient of any quantum theory---its quantum state.  In part, this has been because until recently \cite{Agarwal:2012mq,Collins:2013kqa} a simple language for describing a general state in an inflationary background, aside from particular examples and Bogoliubov transforms of a vacuum state, was largely lacking.  It is also typically assumed that the natural state during inflation is the Bunch-Davies vacuum.  While observations have largely borne out this expectation, it would be interesting to explore both large and small departures from a Bunch-Davies state.  

One observation that simple inflationary models do not adequately explain is the apparent lack of statistical isotropy in some of the low multipole moments in the two-point correlation functions of the CMB \cite{Ade:2015hxq,Schwarz:2015cma}.  In particular, the correlations on the largest angular scales appear anomalously low, and some of the lowest multipole moments appear to be aligned with each other.  Many attempts \cite{Soda:2012zm,Watanabe:2010bu,Chen:2014eua} have been made to explain the origins of these features, usually by including operators that explicitly break the rotational invariance of the standard inflationary picture---for example by selecting a preferred direction or by coupling the scalar and tensor fluctuations directly.  Perhaps the most efficient method for exploring the vast space of Lagrangians for generating these anomalies is through an effective theory treatment \cite{Cheung:2007st,Weinberg:2008hq,Bartolo:2015qvr}.

There are some distinct advantages in considering a more general initial state in inflation, as opposed to modifying the Lagrangian, particularly in the case where one wishes to break space-time symmetries.  The experimental constraints on Lorentz-violating operators constructed from Standard Model fields are generally extremely stringent \cite{Coleman:1998ti}.  Once a fundamental space-time symmetry is explicitly broken, it is hard {\it not\/} to generate relevant Lorentz-violating operators.  One is then compelled to explain why such effects only occurred in the very early universe and why they do not affect the lower-energy fields today.  {\it States\/} that do not respect the underlying symmetries of an action, in contrast, are very easy to generate.  Additionally, when we consider a standard inflationary model in a more general state, Weinberg's theorem \cite{Weinberg:2003sw} for the existence of a conserved adiabatic mode still holds intact since the dynamical degrees of freedom have not changed.

Choosing a particular state leaves open the question of how the universe happened to find itself in that state in the first place.  The approach of this article is that this state is to be understood from a `bottom-up' perspective.  If we parametrise the state in a suitably general way, we can learn what constraints observations place on the state.  This knowledge can then be used to guide what more fundamental, `top-down' dynamics ought to be chosen to lead to an acceptable picture.

In this article we illustrate this idea through a particular example.  We explore how an initial entanglement between the scalar and tensor modes during inflation alters the two-point angular correlation functions of the CMB.  To isolate what is distinctive of this entanglement, we consider a familiar slow-roll theory with a single scalar field, leaving out any structure in the state beyond this entanglement.  The basic theory and description of how the state is fixed appears in the next section.  Then in section 3 we derive the propagators for this theory.  When the initial state has only {\it Gaussian\/} structures in the scalar and tensor modes, their propagators can be solved exactly.  These propagators can then be used to describe how the initial mixing between the two modes evolves to the end of inflation.  In section 4 we show how these new features in the power spectra of the primordial fluctuations lead to new structures in the angular correlators of the CMB.  One distinctive signature that is absent from the standard inflationary picture starting in a Bunch-Davies state is the appearance of nonvanishing off-diagonal angular correlations in the CMB.  In addition, a fairly typical, though not absolutely necessary, feature of these power spectra is the appearance of small oscillations above the Bunch-Davies prediction.

\section{An entangled initial state} 

\noindent 
Let us start by considering a standard class of inflationary models where the expansion is generated by a single scalar field, $\phi(t,\vec x)$.\footnote{We shall be largely following the conventions and notation used in \cite{Maldacena:2002vr,Collins:2011mz}.}  The dynamics of this field and the metric, $g_{\mu\nu}(t,\vec x)$, are determined by an action of the form,
\begin{equation}
S = \int d^4x\, \sqrt{-g}\, \bigl\{ 
{\textstyle{1\over 2}} M_{\rm pl}^2R 
+ {\textstyle{1\over 2}} g^{\mu\nu}\partial_\mu\phi\partial_\nu\phi 
- V(\phi) \bigr\} .
\end{equation}
The solutions to the equations of motion derived from this action are written as a purely time-dependent, classical background, 
\begin{equation}
\phi(t)
\quad\hbox{and}\quad 
ds^2 = dt^2 - e^{2\rho(t)}\, \delta_{ij}\, dx^idx^j ,
\end{equation}
together with the quantum fluctuations about it.  Because we shall be choosing coordinates where they are equivalent, we do not need to distinguish between the field and its classical expectation value.  The detailed form of the potential $V(\phi)$ that appears in the action is also not important here---it is sufficient that it satisfy a pair of slow-roll conditions expressed in terms of two dimensionless parameters $\epsilon$ and $\delta$ which are defined by 
\begin{equation}
\epsilon \equiv {d\over dt}{1\over H} 
= - {\ddot\rho\over\dot\rho^2} \ll 1 ,
\qquad
\delta \equiv {1\over H}{\ddot\phi\over\dot\phi} 
= - {\ddot\phi\over\dot\rho\dot\phi} \ll 1.
\end{equation}
Here $H=\dot\rho$ corresponds to the Hubble scale, the natural energy scale associated with the inflationary expansion.  

When describing small fluctuations about this classical background, we are free to choose our coordinates however we wish as long as we do not alter the simple structure of the background.  Since the background has already singled out the time coordinate, it is convenient to write the general metric, now including the fluctuations, in the form 
\begin{equation}
ds^2 = (N^2-h_{ij}N^iN^j)\, dt^2 - 2h_{ij}N^i\, dtdx^j - h_{ij}\, dx^idx^j . 
\end{equation}
We can use our freedom to choose our coordinates to remove any fluctuations from the scalar field, $\phi(t,\vec x)\equiv\phi(t)$, and also to write the spatial part of the metric as 
\begin{equation}
h_{ij}(t,\vec x) = e^{2\rho(t)+\zeta(t,\vec x)}\, \bigl[ \delta_{ij}+\gamma_{ij}(t,\vec x) \bigr] . 
\end{equation}
Here $\zeta(t,\vec x)$ and $\gamma_{ij}(t,\vec x)$ transform as scalar and tensor fields under the unbroken spatial symmetries of the background metric.  The remaining scalar degrees of freedom in the metric are then fixed by imposing the equations generated by varying $N$ and $N^i$ in the action.  $N$ and $N^i$ turn out to be nondynamical; solving their constraint equations determines their corresponding parts of the metric in terms of the scalar field $\zeta(t,\vec x)$.  To leading order, the result is
\begin{equation}
N(t,\vec x) = 1 + {\dot\zeta\over\dot\rho},
\qquad
N^i(t,\vec x) = \delta^{ij} \partial_j \biggl[
- {e^{-2\rho}\over\dot\rho}\zeta + {1\over 2} {\dot\phi^2\over\dot\rho^2} \partial^{-2}\dot\zeta 
\biggr] .
\end{equation}
The operator $\partial^{-2}$ appearing in the latter expression is the inverse spatial Laplacian, which is straightforwardly defined in terms of the spatial Fourier modes of $\zeta(t,\vec x)$.  Once these constraints have been imposed and the action has been expanded to quadratic order in the scalar and tensor fluctuations, the following leading structure emerges, 
\begin{equation}
S^{(2)} = \int dt d^3\vec x\, {\dot\phi^2\over\dot\rho^2} e^{3\rho} 
\Bigl\{ {\textstyle{1\over 2}}\dot\zeta^2 
- {\textstyle{1\over 2}} e^{-2\rho}\vec\nabla\zeta\cdot\vec\nabla\zeta \Bigr\} 
+ \int dt\, d\vec x\, e^{3\rho}
\Bigl\{ {\textstyle{1\over 8}}\dot\gamma_{ij}\dot\gamma^{ij}
- {\textstyle{1\over 8}}e^{-2\rho}\partial_k\gamma_{ij}\partial^k\gamma^{ij}
\Bigr\} .
\end{equation}

In what follows, it will be much more convenient to work in a momentum representation for the spatial dimensions, so we Fourier transform both of the fields.  The symmetric tensor $\gamma_{ij}(t,\vec x)$ is transverse and traceless, and a standard convention for choosing its two independent components is to define $h_+(t,\vec x)$ and $h_\times(t,\vec x)$.  For instance, for a gravity wave propagating in the $\hat z$ direction these two polarisations correspond to its $\gamma_{11}=-\gamma_{22}$ and $\gamma_{12}=\gamma_{21}$ components,
\begin{equation}
\gamma_{ij} = \begin{pmatrix}
h_+ &h_\times &0\\  
h_\times &-h_+ &0\\  0 &0 &0\
\end{pmatrix} .
\end{equation}
More generally, labelling the two independent polarisations with $\lambda$ which is either $+$ or $\times$, we expand the tensor fluctuations as
\begin{equation}
\gamma_{ij}(t,\vec x) = \sum_{\lambda=+,\times} 
\int {d^3\vec k\over (2\pi)^3}\, e^{i\vec k\cdot\vec x}\; 
e_{ij}^\lambda(\hat k)\; h_{\vec k}^\lambda(t) , 
\end{equation}
where the polarisation tensors $e_{ij}^\lambda(\hat k)$ also satisfy the transverse and traceless conditions, $k^i\, e_{ij}^\lambda(\hat k)=0$ and $\delta^{ij}\, e_{ij}^\lambda(\hat k)=0$.  The polarisations are also orthogonal to each other and normalised with a conventional factor of $2$, since each appears twice in $\gamma_{ij}$,
\begin{equation}
e_{ij}^\lambda(\hat k) \bigl( e^{\lambda' ij}(\hat k) \bigr)^*
= e_{ij}^\lambda(\hat k) e^{\lambda'ij}(-\hat k) 
= 2\delta^{\lambda\lambda'} . 
\end{equation}
If we similarly expand the scalar field, 
\begin{equation}
\zeta(t,\vec x) = \int {d^3\vec k\over (2\pi)^3}\, e^{i\vec k\cdot\vec x}\; 
\zeta_{\vec k}^\lambda(t) , 
\end{equation}
then our earlier quadratic action becomes
\begin{eqnarray}
S^{(2)} &=& 
\int dt \int {d^3\vec k\over (2\pi)^3}\, {\dot\phi^2\over\dot\rho^2} e^{3\rho} 
\Bigl\{ {\textstyle{1\over 2}}\dot\zeta_{-\vec k}\dot\zeta_{\vec k} 
- {\textstyle{1\over 2}} e^{-2\rho} k^2 \zeta_{-\vec k}\zeta_{\vec k} \Bigr\} 
\nonumber \\
&&
+\; \sum_{\lambda=+,\times} 
\int dt \int {d^3\vec k\over (2\pi)^3}\, e^{3\rho}
\Bigl\{ {\textstyle{1\over 8}} \dot h^\lambda_{-\vec k}\dot h^\lambda_{\vec k} 
- {\textstyle{1\over 8}}e^{-2\rho} k^2 h^\lambda_{-\vec k} h^\lambda_{\vec k}
\Bigr\} . 
%\nonumber
\end{eqnarray}

So far we have concentrated on the dynamical side of the theory---this action determines how things evolve.  We next choose an appropriate state in which to evaluate operators.  Here we should like to study what happens when the initial state includes some entanglement between the scalar and tensor modes.  The goal is to learn the characteristic or distinctive signatures in the angular power spectra of the microwave background that such an entanglement would produce.  As long as we are only working to quadratic order, describing the state formally is completely straightforward.  The theory is free, so the Hilbert space for the fluctuations is given by the tensor product of the Hilbert spaces for the scalar and tensor fields, ${\cal H} = {\cal H}_\zeta\otimes{\cal H}_{h_{+,\times}}$, and---depending on the picture in which we are working---either the operators or the states evolve freely from the initial state in which it has started.

The state that typically is used is an asymptotically defined vacuum state,
\begin{equation}
|\Omega\rangle = |0(-\infty)\rangle_\zeta\otimes|0(-\infty)\rangle_h .
\end{equation}
What is meant by this notation is that $|0(-\infty)\rangle$ corresponds to the state that matches with the free Minkowski space vacuum in the asymptotic past, $t\to -\infty$, when all of the Fourier modes $k$ have been blue-shifted to scales where the background curvature is negligible, $k\gg H$.  Of course we are not obliged to consider only this state.  In much the same way that we could explore other models of inflation by including additional fields or operators in the action, we can similarly examine the familiar dynamics of a minimal inflationary model in more complicated quantum states.  Or working still more generally we could consider an arbitrary inflationary model in an arbitrary state.

The way to establish an initial state, defined at a time $t_0$, is through an initial density matrix, $\rho(t_0)\equiv\rho_0$.  To solve the horizon problem, we are obliged to assume that $t_0$ is sufficiently early that the fluctuations responsible for the largest of structures observed today were within the Hubble horizon at this time; but beyond this requirement, $t_0$ is otherwise unconstrained.  For example, $t_0$ does not need to correspond to the beginning of the inflationary expansion; it is only the starting point for our time-evolution.

Let us write the initial density matrix in the form of an action \cite{Agarwal:2012mq,Collins:2013kqa},
\begin{equation}
\rho_0 = e^{iS_0[\zeta(t_0,\vec x),h_{+,\times}(t_0,\vec x)]} .
\end{equation}
The $S_0$ here is integrated over only the spatial coordinates along the initial time hypersurface at $t=t_0$, and all of the fields in $S_0$ are evaluated at $t_0$ as shown.  Once we have chosen a state, the expectation value of an operator ${\cal O}(t)$ is found by evolving its trace with this density matrix forward to the time $t$.  In the interaction picture, this trace would be for example
\begin{equation}
\langle {\cal O}(t)\rangle 
\equiv {\rm tr}\, \bigl[ {\cal O}_I(t) \rho_I(t) \bigr]
= {\rm tr}\, \bigl[ U_I^\dagger(t,t_0) {\cal O}_I(t) U_I(t,t_0)\rho_0 \bigr] ,
\end{equation}
where $U_I(t,t_0)$ is the familiar time-evolution operator constructed from the interacting part of the Hamiltonian, $H_I$,
\begin{equation}
U_I(t,t_0) = T e^{-i\int_{t_0}^t dt'\, H_I(t')} . 
\end{equation}
If the state contains structures that are higher order than what can be expressed quadratically in the fields, these could be included in $H_I$ as well.  The expectation value of the operator is then 
\begin{equation}
\langle {\cal O}(t)\rangle 
= {\rm tr}\, \bigl[ \bigl( T e^{-i\int_{t_0}^t dt'\, H_I(t')}\bigr)^\dagger {\cal O}_I(t) \bigl( T e^{-i\int_{t_0}^t dt'\, H_I(t')}\bigr) e^{iS_0[\zeta(t_0,\vec x),h_{+,\times}(t_0,\vec x)]} \bigr] .
\end{equation}
While this expression for $\langle {\cal O}(t)\rangle$ could be used just as it is written \cite{Weinberg:2005vy}, it is more useful to introduce `$+$' and `$-$' versions of the fields to write it in terms of a single time-ordered operator,
$$
\langle {\cal O}(t)\rangle 
= {\rm tr}\, \Bigl[ T \Bigl( {\cal O}^+_I(t) e^{-i\int_{t_0}^t dt'\, 
[H_I[\zeta^+(t',\vec x),h^+_\lambda(t',\vec x)] 
- H_I[\zeta^-(t',\vec x),h^-_\lambda(t',\vec x)]]} e^{iS_0[\zeta^\pm(t_0,\vec x),h^\pm_\lambda(t_0,\vec x)]} \Bigr) \Bigr] .
$$
The `$+$' is meant to indicate a field that appeared in the $U_I(t,t_0)$ operator, whereas a `$-$' indicates one that appeared in the $U_I^\dagger(t,t_0)$ operator.  If this expression is to reproduce the original one, the convention is that the `$-$' fields always occur {\it later\/} the `$+$' fields in the sense that the time-ordering places `$-$' fields to the left of the `$+$' fields.  The `$-$' fields are also anti-time-ordered because the Hermitian conjugation reverses the order of the operators.

Here we are analysing just a quadratic theory, so the problem of evolving the expectation value of an operator reduces to the problem of solving the free time-dependence of the fields and leaving the density matrix in its initial state.  That is, the problem simplifies to solving a free theory in the Heisenberg picture, 
\begin{equation}
\langle {\cal O}(t)\rangle \equiv {\rm tr}\, \bigl[ {\cal O}_H(t) \rho_0 \bigr] .
\end{equation}
We retain the $\pm$ notation nonetheless when writing the density matrix of the initial state. 

Now let us make a few simplifications.  Since we have only expanded the dynamical part of the action to quadratic order in the scalar and tensor fields, we shall do the same for $S_0$.  Furthermore, since here we wish to investigate the signatures that would be associated specifically with some entanglement between the scalar and tensor fluctuations, as opposed to considering the most general initial state, we shall not introduce any {\it separate\/} structures for either of the fields.  Such structures would be represented by terms in $S_0$ that are either quadratic in $\zeta(t_0\vec x)$ {\it or\/} quadratic in $h_{+,\times}(t_0,\vec x)$.  These restrictions leave us with an initial action containing terms that are linear in both fields at once, 
\begin{eqnarray}
S_0 &=& - \int d^3\vec x\, d^3\vec y\, \Bigl\{
A^{ij}(\vec x-\vec y)\, \zeta^+(t_0,\vec x)\gamma_{ij}^+(t_0,\vec y)
- A^{ij*}(\vec x-\vec y)\, 
\zeta^-(t_0,\vec x)\gamma_{ij}^-(t_0,\vec y)
\nonumber \\
&&\qquad\qquad\  
+\,\, iB^{ij}(\vec x-\vec y)\, 
\zeta^-(t_0,\vec x)\gamma_{ij}^+(t_0,\vec y)
+ i B^{ij*}(\vec x-\vec y)\, 
\zeta^+(t_0,\vec x)\gamma_{ij}^-(t_0,\vec y)
\Bigr\} . \qquad
%\nonumber 
\end{eqnarray}
The initial density matrix is Hermitian, $\rho_0^\dagger = \rho_0$ or $S_0^\dagger = -S_0$, which in turn implies that $A^{ij}(\vec x-\vec y)$ is a complex function while $B^{ij}(\vec x-\vec y)$ is purely real.  By expanding the structure functions as 
\begin{equation}
A^{ij}(\vec x-\vec y) = \int {d^3\vec k\over (2\pi)^3}\, 
e^{i\vec k\cdot(\vec x-\vec y)} A^{ij}_{\vec k} 
\quad\hbox{and}\quad
B^{ij}(\vec x-\vec y) = \int {d^3\vec k\over (2\pi)^3}\, 
e^{i\vec k\cdot(\vec x-\vec y)} B^{ij}_{\vec k} , 
\end{equation}
we can write this action as 
\begin{eqnarray}
S_0 &=& - \sum_{\lambda=+,\times} 
\int {d^3\vec k\over (2\pi)^3}\, \Bigl\{
\bigl[ A^{ij}_{\vec k}e_{ij}^\lambda(\hat k)\bigr]\, 
\zeta^+_{-\vec k}(t_0) h_{\vec k}^{\lambda+}(t_0)
- \bigl[ A^{ij*}_{\vec k}e_{ij}^\lambda(-\hat k)\bigr]\, 
\zeta^-_{-\vec k}(t_0) h_{\vec k}^{\lambda-}(t_0)
\nonumber \\
&&\qquad\qquad\qquad\   
+\,\, i \bigl[ B^{ij}_{\vec k}e_{ij}^\lambda(\hat k)\bigr]\, 
\zeta^-_{-\vec k}(t_0) h_{\vec k}^{\lambda+}(t_0)
+ i \bigl[ B^{ij*}_{\vec k}e_{ij}^\lambda(-\hat k)\bigr]\, 
\zeta^+_{-\vec k}(t_0) h_{\vec k}^{\lambda-}(t_0)
\Bigr\} . \qquad\quad
%\nonumber 
\end{eqnarray}
When the particular structure functions $B_{\vec k}^{ij}$ are nonvanishing, the corresponding initial state cannot be obtained through a Bogoliubov transformation of the Bunch-Davies vacuum state, even if we allow transformations in a more general sense that mix the momentum modes of the scalar and tensor fields.

Notice that $S_0$ is (i) not local in space and (ii) not rotationally invariant.  The non-locality occurs here for the reason that there can be structures or correlations already present in the state at $t_0$.  We have chosen the structures that describe the initial state, $A^{ij}(\vec x-\vec y)$ and $B^{ij}(\vec x-\vec y)$, to be invariant under translations, but they will break the rotational invariance of the dynamical part of the action.  The violation of the rotational invariance can enter through the state in two ways.  There is the inevitable breaking that occurs simply because an operator that couples scalar and tensor fields is not rotationally invariant.  The second possibility is that the structure functions themselves might contain some further directional dependence.  To keep the analysis relatively simple, we only include the former by assuming that $A^{ij}(\vec x-\vec y)$ and $B^{ij}(\vec x-\vec y)$ depend on $|\!|\vec x-\vec y|\!|$ but not the direction of $(\vec x-\vec y)$.  Because the matrix containing the tensor fluctuations $\gamma_{ij}^\pm(t,\vec x)$ is symmetric, transverse, and traceless, only the corresponding components of the structure matrices for the initial state will contribute; any other components are projected out by the contraction between $A^{ij}$ and $B^{ij}$ with $\gamma_{ij}^\pm$.  Let us write these components of the structure functions in the same basis that we used to expand the tensor fields, 
$$
A^{ij}_{\vec k} = {\textstyle{1\over 2}}\sum_{\lambda=+,\times} 
e^{\lambda ij}(-\hat k)\; A^\lambda_k + \cdots, \qquad
A^{ij*}_{\vec k} = {\textstyle{1\over 2}}\sum_{\lambda=+,\times} 
e^{\lambda ij}(\hat k)\; A^{\lambda *}_k + \cdots, 
$$
and 
$$
B^{ij}_{\vec k} = {\textstyle{1\over 2}}\sum_{\lambda=+,\times} 
e^{\lambda ij}(-\hat k)\; B^\lambda_k, + \cdots\qquad
B^{ij*}_{\vec k} = {\textstyle{1\over 2}}\sum_{\lambda=+,\times} 
e^{\lambda ij}(\hat k)\; B^\lambda_k + \cdots. 
$$
The rotational invariance of $|\!|\vec x-\vec y|\!|$ is reflected in the fact that $A^\lambda_k$ and $B^\lambda_k$ depend only on the magnitude of $k\equiv |\!|\vec k|\!|$.

In this scenario, the action associated with the initial state becomes 
\begin{eqnarray}
S_0 &=& - \sum_{\lambda=+,\times} 
\int {d^3\vec k\over (2\pi)^3}\, \Bigl\{
A^\lambda_k \zeta^+_{-\vec k}(t_0) h_{\vec k}^{\lambda+}(t_0)
- A^{\lambda *}_k \zeta^-_{-\vec k}(t_0) h_{\vec k}^{\lambda-}(t_0)
\nonumber \\
&&\qquad\qquad\qquad\   
+\,\, i B^\lambda_k \zeta^-_{-\vec k}(t_0) h_{\vec k}^{\lambda+}(t_0)
+ iB^\lambda_k \zeta^+_{-\vec k}(t_0) h_{\vec k}^{\lambda-}(t_0)
\Bigr\} . 
%\nonumber 
\end{eqnarray}
Although we have chosen a comparatively simple state, where the fields can exchange angular momentum with the background described by the functions, $A_k^\lambda$ and $B_k^\lambda$, it is important to remember that this formalism can be applied to states that explicitly break the rotational invariance or other symmetries as well.  One phenomenologically persuasive reason that we might wish to break these symmetries is that they might be used to generate the observed asymmetries at large scales in the microwave background \cite{Ade:2015hxq,Schwarz:2015cma}.

The free action is all that is needed to compute the two-point functions for the primordial fluctuations at leading order.  Even so, the propagators for this theory are considerably more complicated than those evaluated in the pure vacuum state, or even in a merely unentangled state.  In addition to the scalar propagator, which we shall write as 
\begin{equation}
{\rm tr}\, \bigl[ T\bigl(\zeta^\pm(t,\vec x)\zeta^\pm(t',\vec y) \rho_0 \bigr) \bigr] \equiv \int {d^3\vec k\over (2\pi)^3}\, 
e^{i\vec k\cdot(\vec x-\vec y)} {\bf G}_k^{\pm\pm}(t,t') ,
\end{equation}
and those for the separate polarisations of the tensor fields, there will also be mixings between the scalar and tensor fields,
\begin{equation}
{\rm tr}\, \bigl[ T\bigl(\zeta^\pm(t,\vec x)h_\lambda^\pm(t',\vec y) \rho_0 \bigr) \bigr] \equiv \int {d^3\vec k\over (2\pi)^3}\, 
e^{i\vec k\cdot(\vec x-\vec y)} {\bf M}_{\lambda,k}^{\pm\pm}(t,t') ,
\end{equation}
and through such mixings, the polarisations of the tensor fields also become mixed,
\begin{equation}
{\rm tr}\, \bigl[ T\bigl(h_\lambda^\pm(t,\vec x)h_{\lambda'}^\pm(t',\vec y) \rho_0 \bigr) \bigr] \equiv \int {d^3\vec k\over (2\pi)^3}\, 
e^{i\vec k\cdot(\vec x-\vec y)} {\bf H}_{\lambda\lambda',k}^{\pm\pm}(t,t') .
\end{equation}
The superscripts on the fields mean that we have four types of each propagator, corresponding to the four choices for the $\pm\pm$ indices introduced earlier.  For these entangled states, there are thus both cross-correlations between the scalar and tensor fields and mixings between different polarisations.  Neither of these effects occurs for correlators evaluated in the standard vacuum state.

Although the primordial power spectra correspond to setting $t'=t$ in the propagators of the entangled state, we have described how to treat a fully interacting theory since the formalism that we have developed in this section can also be used to explore non-Gaussianities or corrections due to cubic and higher order interactions in the full action, $S+S_0$.

\section{Propagators and two-point functions} 

\noindent 
So to find the two-point functions, what we need to do is to solve for the propagators in an entangled state.  A straightforward method for doing this is to regard $S$ temporarily as the free action and to treat $S_0$, the part associated with the entangled initial state, as an `interaction'.  Later we shall consider $S_0$ to be `small', in the sense of its being a small departure from the standard vacuum state; but at this stage we should like to set up the general calculation and not to restrict to the perturbative case from the start.  In analysing the theory described by $S+S_0$ in this interaction picture, we still need to choose an appropriate state from which to build up the state described by $\rho_0$.  A natural and convenient, although by no means obligatory, choice is the usual Bunch-Davies vacuum state.  From this perspective, $S_0$ describes a particular set of excitations with respect to the standard vacuum.

The propagator for the entangled state equals the sum of all insertions of the `interactions' from $S_0$ connected by Bunch-Davies propagators.  We shall denote the Bunch-Davies propagators for the scalar and tensor fields by $G_k^{\pm\pm}(t,t')$ and $H_{\lambda\lambda,k}^{\pm\pm}(t,t')$.  These propagators can in turn be expressed in terms of Wightman functions, which for the scalar fields are
\begin{eqnarray}
\langle\Omega|\zeta(t,\vec x)\zeta(t',\vec y)|\Omega\rangle 
&\equiv& \int {d^3\vec k\over (2\pi)^3}\, e^{i\vec k\cdot(\vec x-\vec y)} 
G^>_k(t,t') 
\nonumber \\
\langle\Omega|\zeta(t',\vec y)\zeta(t,\vec x)|\Omega\rangle 
&\equiv& \int {d^3\vec k\over (2\pi)^3}\, e^{i\vec k\cdot(\vec x-\vec y)} 
G^<_k(t,t') , 
%\nonumber 
\end{eqnarray}
with similar Wightman functions defined for the polarisations of the tensor fields, $H^{>,<}_{\lambda\lambda,k}(t,t')$.  At equal times they are equal, so $G^>_k(t,t) = G^<_k(t,t)$.  The four propagators for the scalar field are distinguished from each other by the choices for the $\pm$ indices,
\begin{eqnarray}
G_k^{++}(t,t') &=& \Theta(t-t')\, G_k^>(t,t') + \Theta(t'-t)\, G_k^<(t,t')
\nonumber \\
G_k^{+-}(t,t') &=& G_k^<(t,t')
\nonumber \\
G_k^{-+}(t,t') &=& G_k^>(t,t')
\nonumber \\
G_k^{--}(t,t') &=& \Theta(t'-t)\, G_k^>(t,t') + \Theta(t-t')\, G_k^<(t,t') , 
%\nonumber 
\end{eqnarray}
where the time-ordering corresponds to what we described earlier for the $\pm$ fields.  A little later we shall provide explicit expressions for the Bunch-Davies Wightman functions, but for now we write the entangled propagators in terms of $G_k^{>,<}(t,t')$ and $H_{\lambda\lambda,k}^{>,<}(t,t')$.

With the Wightman functions defined thus, we are ready to compute the entangled propagator by summing the corrections due to the interactions at $t_0$.  The zeroth order contribution, provided there is one, is just the Bunch-Davies propagator itself.  But because the terms in $S_0$ connect different fields, the internal lines of the higher-order corrections can be either scalar or tensor vacuum propagators.  The interaction in this picture only occurs at $t_0$; from the perspective of the time-ordering of the `+' fields this is the earliest possible time while for the `$-$' fields this is the latest possible time.  So the contributions from the  higher corrections are universal in the sense that the entangled propagators can always be written as the Bunch-Davies propagator, which does depend on the $\pm$ indices, plus a term that does not,\footnote{There are additional corrections to the propagator at $t_0$ that are required for the consistency of the Green's functions.  These vanish unless {\it both\/} of the times occurring in the propagator are the initial time {\it and\/} a time derivative is present.  These corrections are not required for computing the power spectra.} 
\begin{eqnarray}
{\bf G}_k^{\pm\pm}(t,t')  &=& 
G^{\pm\pm}_k(t,t') + \Delta G_k(t,t') 
+ \hbox{terms that vanish for $t,t'>t_0$}
\nonumber \\
{\bf M}_{\lambda,k}^{\pm\pm}(t,t')  &=& 
\Delta M_{\lambda,k}(t,t') 
+ \hbox{terms that vanish for $t,t'>t_0$}
\nonumber \\
{\bf H}_{\lambda\lambda',k}^{\pm\pm}(t,t')  &=&
H_{\lambda\lambda',k}^{\pm\pm}(t,t') 
+ \Delta H_{\lambda\lambda',k}(t,t') 
+ \hbox{terms that vanish for $t,t'>t_0$} . 
\nonumber 
\end{eqnarray}
As an example, the term $\Delta G(t,\vec x;t',\vec y)$ is given by the sum of the following infinite family of graphs,
\begin{equation}
\includegraphics[scale=1]{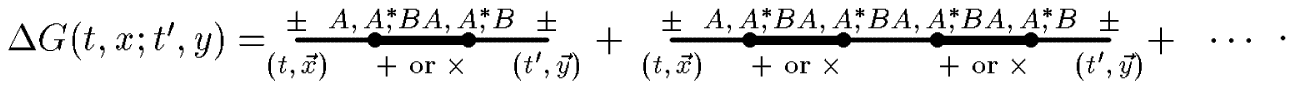}
\end{equation}
In this equation, a thin line indicates a scalar propagator while a thick line represents one of the tensor propagators.  The sum of this series is 
\begin{eqnarray}
\Delta G_k(t,t') &=&
{H_k^>(t_0,t_0)\over 1 + G^>_k(t_0,t_0)H^>_k(t_0,t_0)[(A_k^+-A_k^{+*}+2iB_k^+)^2 + (A_k^\times-A_k^{\times*}+2iB_k^\times)^2]} 
\nonumber \\
&&\times
\Bigl\{ - G_k^>(t,t_0)G_k^>(t',t_0) 
\Bigl[ (A^+_k+iB^+_k)^2 + (A^\times_k+iB^\times_k)^2 \Bigr] 
\nonumber \\
&&\quad\ \ 
-\,\, G_k^<(t,t_0)G_k^<(t',t_0) 
\Bigl[ (A^{+*}_k-iB^+_k)^2 + (A^{\times*}_k-iB^\times_k)^2 \Bigr]
\nonumber \\
&&\quad\ \ 
+\,\, \bigl[ G_k^>(t,t_0)G_k^<(t',t_0) + G_k^<(t,t_0)G_k^>(t',t_0) \bigr]  
\Bigl[ |A^+_k+iB^+_k|^2 + |A^\times_k+iB^\times_k|^2 \Bigr] 
\nonumber \\
&&\quad\ \ 
- \bigl[ G_k^>(t,t_0)-G_k^<(t,t_0) \bigr] \bigl[ G_k^>(t',t_0) - G_k^<(t',t_0) \bigr] G^>_k(t_0,t_0) H^>_k(t_0,t_0) 
\nonumber \\
&&\qquad\times
\Bigl[ A^+_k A^{\times*}_k - A^{+*}_k A^\times_k 
+ iB^+_k(A^\times_k+A^{\times*}_k) - iB^\times_k(A^+_k+A^{+*}_k) \Bigr]^2
\Bigr\} .
%\nonumber 
\end{eqnarray}

We could perform the analogous calculation for the tensor propagator, summing again over an infinite series of insertions of the operators associated with the initial state.  However, since we have already gone to the trouble of computing the scalar propagator for the entangled state, we can circumvent this laborious calculation by attaching Bunch-Davies propagators to its ends to generate the entangled-state propagator for the tensor fields directly,
\begin{eqnarray}
\Delta H_{\lambda\lambda',k}(t,t')
&=&
{G^>_k(t_0,t_0)\over 1 + G^>_k(t_0,t_0)H^>_k(t_0,t_0)[(A_k^+-A_k^{+*}+2iB_k^+)^2 + (A_k^\times-A_k^{\times*}+2iB_k^\times)^2]} 
\nonumber \\
&&\times
\Bigl\{ - H_k^>(t,t_0)H_k^>(t',t_0) 
\bigl[ A_k^\lambda+iB_k^\lambda \bigr] \bigl[ A_k^{\lambda'}+iB_k^{\lambda'} \bigr] 
\nonumber \\
&&\quad\ 
- H_k^<(t,t_0)H_k^<(t',t_0) 
\bigl[ A_k^{\lambda*}-iB_k^\lambda \bigr] \bigl[ A_k^{\lambda'*}-iB_k^{\lambda'} \bigr] 
\nonumber \\
&&\quad\
+ H_k^>(t,t_0)H_k^<(t',t_0) 
\bigl[ A_k^\lambda+iB_k^\lambda \bigr] \bigl[ A_k^{\lambda'*}-iB_k^{\lambda'} \bigr] 
\nonumber \\
&&\quad\ 
+ H_k^<(t,t_0)H_k^>(t',t_0) 
\bigl[ A_k^{\lambda*}-iB_k^\lambda \bigr] \bigl[ A_k^{\lambda'}+iB_k^{\lambda'} \bigr] 
\Bigr\} . 
%\nonumber 
\end{eqnarray}
Notice that while the polarisations were uncoupled in the free Bunch-Davies state, the entanglement has caused them to become mixed together---that is, $\Delta H_{\lambda\lambda',k}(t,t')$ does not vanish if $\lambda\not=\lambda'$.  We can connect a $+$ polarised fluctuation to a $\times$ polarised fluctuation through a scalar propagator.  For example, the leading graph that couples the different polarisations is 
$$
\includegraphics[scale=1]{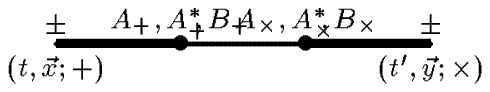}
$$
And of course, nothing prevents a direct conversion between the scalar and tensor fields in this theory, 
\begin{eqnarray}
{\bf M}_{\lambda,k}^{\pm\pm}(t,t') &=&
{1\over 1 + G^>_k(t_0,t_0)H^>_k(t_0,t_0)[(A_k^+-A_k^{+*}+2iB_k^+)^2 + (A_k^\times-A_k^{\times*}+2iB_k^\times)^2]} 
\\
&&\times
\Bigl\{ -i G^>_k(t,t_0) \bigl[ A_k^\lambda H_k^>(t',t_0) + iB_k^\lambda H_k^<(t',t_0)\bigr]  
\nonumber \\
&&\quad\ 
+ i G^<_k(t,t_0) \bigl[ A_k^{\lambda*} H_k^<(t',t_0) - iB_k^\lambda H_k^>(t',t_0)\bigr] 
\nonumber \\
&&\quad\ 
+ i G^>_k(t_0,t_0)H^>_k(t_0,t_0) [G^>_k(t,t_0) - G^<_k(t,t_0)]
\bigl[ A_k^\lambda H_k^>(t',t_0) + iB_k^\lambda H_k^<(t',t_0)\bigr]  
\nonumber \\
&&\qquad\times
[(A_k^{+*}-iB_k^+)(A_k^+-A_k^{+*}+2iB_k^+) + (A_k^{\times*}-iB_k^\times)(A_k^\times-A_k^{\times*}+2iB_k^\times)] 
\nonumber \\
&&\quad\ 
- i G^>_k(t_0,t_0)H^>_k(t_0,t_0) [G^>_k(t,t_0) - G^<_k(t,t_0)]
\bigl[ A_k^{\lambda*} H_k^<(t',t_0) - iB_k^\lambda H_k^>(t',t_0)\bigr]  
\nonumber \\
&&\qquad\times
[(A_k^++iB_k^+)(A_k^+-A_k^{+*}+2iB_k^+) + (A_k^\times +iB_k^\times)(A_k^\times-A_k^{\times*}+2iB_k^\times)] 
\Bigr\} , 
\nonumber 
\end{eqnarray}
which has no zeroth-order term because the vacuum propagator does not connect the scalar and tensor modes.

The quantity that appears in the denominator of the prefactor of each propagator is a purely nonpositive function,
\begin{equation}
G^>_k(t_0,t_0)H^>_k(t_0,t_0)[(A_k^+-A_k^{+*}+2iB_k^+)^2 + (A_k^\times-A_k^{\times*}+2iB_k^\times)^2] \le 0 .
\end{equation}
So if we want to avoid having a pole in the propagators, then we should require
\begin{equation}
\bigl| G^>_k(t_0,t_0)H^>_k(t_0,t_0)[(A_k^+-A_k^{+*}+2iB_k^+)^2 + (A_k^\times-A_k^{\times*}+2iB_k^\times)^2] \bigr| < 1 .
\end{equation}
We shall return to this bound later when we have explicit expressions for the Wightman functions.

Once we have derived the propagators, all that is needed to compute the two-point functions is to set $t'=t$, 
\begin{eqnarray}
\langle\zeta_{-\vec k}(t)\zeta_{\vec k}(t)\rangle_\rho &\equiv&
{\bf G}_k(t,t) = G^>_k(t,t) + \Delta G_k(t,t) 
\nonumber \\
\langle\zeta_{-\vec k}(t) h^\lambda_{\vec k}(t)\rangle_\rho &\equiv&
{\bf M}_{\lambda,k}(t,t) = \Delta M_{\lambda,k}(t,t) 
\nonumber \\
\langle h^{\lambda'}_{-\vec k}(t) h^\lambda_{\vec k}(t)\rangle_\rho &\equiv& 
{\bf H}_{\lambda\lambda',k}(t,t) =
H_{\lambda\lambda',k}^>(t,t) + \Delta H_{\lambda\lambda',k}(t,t) . 
%\nonumber 
\end{eqnarray}
While we did not need the full expressions for the propagators to compute the two-point functions, we have derived them here nonetheless since we would need these propagators to study the influence of cubic and higher order interactions in both the `dynamical' part of the action, $S$, and the `initial' part of the action, $S_0$.  Such interactions would contribute to the three-point function---and we could similarly add higher-order operators to the initial state too---as well as to loop corrections to the two-point function.

Thus far we have been writing the propagators in terms of the Wightman functions to keep the expressions relatively compact and fairly general.  These functions can also be expressed in terms of the mode functions, $\zeta_k(t)$ and $h_k(t)$, associated with the expansions of the fields in creation and annihilation operators,\footnote{Note the subtle difference in the notation:  the mode function $\zeta_k(t)$ is an ordinary complex function of $k$ and $t$ whereas the $\zeta_{\vec k}(t)$ written earlier is an operator.}
\begin{eqnarray}
\zeta(t,\vec x)  &=& \int {d^3\vec k\over (2\pi)^3}\, 
\bigl\{ e^{i\vec k\cdot\vec x} \zeta_k(t) a_{\vec k} 
+ e^{-i\vec k\cdot\vec x} \zeta_k^*(t) a_{\vec k}^\dagger \bigr\}
\nonumber \\
h^\lambda(t,\vec x) &=& \int {d^3\vec k\over (2\pi)^3}\, 
\bigl\{ e^{i\vec k\cdot\vec x} h_k(t) a^\lambda_{\vec k} 
+ e^{-i\vec k\cdot\vec x} h_k^*(t) a_{\vec k}^{\lambda\dagger} \bigr\} .
%\nonumber
\end{eqnarray}
When the inflationary background is in the slow-roll limit, these mode functions can be approximated by their de Sitter forms.  Applying the prescriptions that these functions assume the forms appropriate for the Minkowski vacuum in the limit $t\to -\infty$, or equivalently when $k\to\infty$, and that they have a canonical normalisation, they are
\begin{equation}
\zeta_k(t) = {i\over 2\sqrt{\epsilon}} {H\over M_{\rm pl}} 
{(1+ik\eta)\over k^{3/2}} e^{-ik\eta} + \cdots 
\end{equation}
and 
\begin{equation}
h_k(t) = i {H\over M_{\rm pl}} 
{(1+ik\eta)\over k^{3/2}} e^{-ik\eta} + \cdots 
\end{equation}
to leading order in $\epsilon$ and $\delta$.  We have replaced the time coordinate with the conformal time, $d\eta = dt/a(t)$, in these expressions since the functions become more compact when expressed in terms of the conformal time $\eta$.  The Bunch-Davies Wightman functions are then
\begin{equation}
G_k^>(t,t') = \zeta_k(t)\zeta_k^*(t') ,
\qquad
G_k^<(t,t') = \zeta_k^*(t)\zeta_k(t') ,
\end{equation}
and
\begin{equation}
H_k^>(t,t') = h_k(t)h_k^*(t') ,
\qquad
H_k^<(t,t') = h_k^*(t)h_k(t') .
\end{equation}
Since the two-point functions are evaluated at very late times, in the sense that the modes responsible for the fluctuations in the microwave background radiation have been stretched far outside the horizon, we can evaluate them in the $t\to\infty$ or $\eta\to 0$ (or more correctly, $k\eta\to 0$) limits in which
\begin{equation}
\zeta_k(t) = {i\over 2\sqrt{\epsilon}} {H\over M_{\rm pl}} 
{1\over k^{3/2}} + \cdots 
\qquad\hbox{and}\qquad 
h_k(t) = i {H\over M_{\rm pl}} 
{1\over k^{3/2}} + \cdots .
\end{equation}
In contrast, to allow the initial state to have been causally generated, the modes at the initial times should be inside the horizon; {\it minimally\/} this means that $|k\eta_0|>1$.  In the de Sitter limit, $\eta$ depends exponentially on the coordinate $t$, so we could assume that $|k\eta_0|\gg 1$ if we wished.

In terms of the Bunch-Davies mode functions, the two-point function for the scalar field becomes 
\begin{eqnarray}
\langle \zeta_{-\vec k}(t)\zeta_{\vec k}(t)\rangle &=& 
{|\zeta_k(t)|^2\over 1 + |\zeta_k(t_0)|^2 |h_k(t_0)|^2 \bigl\{ 
\bigl[ A_k^+ - A_k^{+*} + 2iB_k^+ \bigr]^2 
+ \bigl[ A_k^\times - A_k^{\times *} + 2iB_k^\times \bigr]^2 
\bigr\} } 
\nonumber \\
&&
\biggl\{ 1 
+ |\zeta_k(t_0)|^2 |h_k(t_0)|^2 
\biggl[ 1 + {\zeta_k^*(t_0)\over\zeta_k(t_0)} \biggr]
\Bigl[ (A^+_k+iB^+_k)^2 + (A^\times_k+iB^\times_k)^2 \Bigr]
\nonumber \\
&&\quad\ \ 
+\, |\zeta_k(t_0)|^2 |h_k(t_0)|^2 
\biggl[ 1 + {\zeta_k(t_0)\over\zeta_k^*(t_0)} \biggr]
\Bigl[ (A^{+*}_k-iB^+_k)^2 + (A^{\times *}_k-iB^\times_k)^2 \Bigr]
\nonumber \\
&&\quad\ \ 
+\, |\zeta_k(t_0)|^4 |h_k(t_0)|^4 
\biggl[ 2 + {\zeta_k^*(t_0)\over\zeta_k(t_0)}
+ {\zeta_k(t_0)\over\zeta_k^*(t_0)} \biggr] 
\nonumber \\
&&\qquad\quad\times 
\Bigl[ A_k^+A_k^{\times *} - A_k^\times A_k^{+*} 
+ iB_k^+(A_k^\times + A_k^{\times *}) - iB_k^\times(A_k^+ + A_k^{+*}) \Bigr]^2 
\biggr\}, \qquad\quad
%\nonumber 
\end{eqnarray}
while for the tensor fields it is 
\begin{eqnarray}
&&\hskip-0.375truein
\langle h^\lambda_{-\vec k}(t)h^{\lambda'}_{\vec k}(t)\rangle 
 \\
&=& 
|h_k(t)|^2
\biggl\{ \delta^{\lambda\lambda'}
+ {|\zeta_k(t_0)|^2 |h_k(t_0)|^2 \over 1 + |\zeta_k(t_0)|^2 |h_k(t_0)|^2 
\bigl\{ 
\bigl[ A_k^+ - A_k^{+*} + 2iB_k^+ \bigr]^2 
+ \bigl[ A_k^\times - A_k^{\times *} + 2iB_k^\times \bigr]^2 \bigr\} } 
\nonumber \\
&&\qquad\qquad\quad\quad\times  
\biggl[ {h_k^*(t_0)\over h_k(t_0)} 
\bigl[ A_k^\lambda +iB_k^\lambda \bigr] \bigl[ A_k^{\lambda'} +iB_k^{\lambda'} \bigr]
+ {h_k(t_0)\over h_k^*(t_0)} \bigl[ A_k^{\lambda*}-iB_k^\lambda \bigr] 
\bigl[ A_k^{\lambda' *}-iB_k^{\lambda'} \bigr]
\nonumber \\
&&\qquad\qquad\quad\qquad 
+\; \bigl[ A_k^\lambda +iB_k^\lambda \bigr] 
\bigl[ A_k^{\lambda' *} -iB_k^{\lambda'} \bigr]
+ \bigl[ A_k^{\lambda*} -iB_k^\lambda \bigr]
\bigl[ A_k^{\lambda'} +iB_k^{\lambda'} \bigr] 
\biggr]
\biggr\} ,
\nonumber 
\end{eqnarray}
and there is a mixed scalar-tensor two-point function as well, 
\begin{eqnarray}
&&\hskip-0.375truein
\langle \zeta_{-\vec k}(t)h^\lambda_{\vec k}(t)\rangle
 \\
&=&
{1\over 1 + |\zeta_k(t_0)|^2 |h_k(t_0)|^2 
[(A_k^+-A_k^{+*}+2iB_k^+)^2 + (A_k^\times-A_k^{\times*}+2iB_k^\times)^2]} 
\nonumber \\
&&\times
\Bigl\{ -i \zeta_k(t)\zeta_k^*(t_0) 
\bigl[ A_k^\lambda h_k(t)h_k^*(t_0) + iB_k^\lambda h_k^*(t)h_k(t_0)\bigr]  
\nonumber \\
&&\quad\ 
+ i \zeta_k^*(t)\zeta_k(t_0) 
\bigl[ A_k^{\lambda*} h_k^*(t)h_k(t_0) - iB_k^\lambda h_k(t)h_k^*(t_0)\bigr] 
\nonumber \\
&&\quad\ 
+ i |\zeta_k(t_0)|^2 |h_k(t_0)|^2 [\zeta_k(t)\zeta_k^*(t_0) - \zeta_k^*(t)\zeta_k(t_0)]
\bigl[ A_k^\lambda h_k(t)h_k^*(t_0) + iB_k^\lambda h_k^*(t)h_k(t_0)\bigr]  
\nonumber \\
&&\qquad\times
[(A_k^{+*}-iB_k^+)(A_k^+-A_k^{+*}+2iB_k^+) + (A_k^{\times*}-iB_k^\times)(A_k^\times-A_k^{\times*}+2iB_k^\times)] 
\nonumber \\
&&\quad\ 
- i |\zeta_k(t_0)|^2 |h_k(t_0)|^2 [\zeta_k(t)\zeta_k^*(t_0) - \zeta_k^*(t)\zeta_k(t_0)]
\bigl[ A_k^{\lambda*} h_k^*(t)h_k(t_0) - iB_k^\lambda h_k(t)h_k^*(t_0)\bigr]  
\nonumber \\
&&\qquad\times
[(A_k^++iB_k^+)(A_k^+-A_k^{+*}+2iB_k^+) + (A_k^\times +iB_k^\times)(A_k^\times-A_k^{\times*}+2iB_k^\times)] 
\Bigr\} .
\nonumber 
\end{eqnarray}
Note that each of these two-point functions is purely real.  One consequence is that 
\begin{equation}
\langle h^+_{-\vec k}(t)h^\times_{\vec k}(t)\rangle
= \langle h^\times_{-\vec k}(t)h^+_{\vec k}(t)\rangle ,
\end{equation}
which will be useful to remember when later we switch to a helicity basis for the polarisations.

The condition that we should impose if we wish to avoid a state-dependent pole in the propagators becomes, 
\begin{equation}
\Bigl| [(A_k^+-A_k^{+*}+2iB_k^+)^2 + (A_k^\times-A_k^{\times*}+2iB_k^\times)^2] \Bigr| 
< 4|\epsilon| {M_{\rm pl}^4\over H^4} {k^6\over (1+k^2\eta_0^2)^2} .
\end{equation}
We can see that as $k\to 0$, the state in this case should be such that ${\rm Im}\,A_k^+\to -B_k^+$ and ${\rm Im}\,A_k^\times\to -B_k^\times$.

\subsection{Rescaling and small departures from the vacuum} 

\noindent
Before going further, we need to understand what is the `natural' scaling of the functions $A_k^{+,\times}$ and $B_k^{+,\times}$ in $k$ and what is meant by a `small' departure from the vacuum state.  It was natural to write these functions $A_k^{+,\times}$ and $B_k^{+,\times}$ as the coefficients of the operators mixing the scalar and tensor structures in the action for the initial state, but they are not dimensionless, but rather have a mass dimension of $3$.  We should like to recast these functions in a dimensionless form to parametrise how the state differs from the vacuum.  Such a description will also be important in the next section when we analyse an example of the characteristic signatures of this scenario for the two-point angular correlators of the CMB.

There are several mass scales associated with inflation---$k$, $H$, $M_{\rm pl}$, and $1/\eta_0$---but a natural choice is to absorb some of the factors that appear in $\zeta_k(t_0)$ and $h_k(t_0)$.  If we express the action associated with the initial state in terms of the Bunch-Davies creation and annihilation operators, we find terms of the form
\begin{eqnarray}
S_0 &=& 
- \int {d^3\vec k\over (2\pi)^3}\, \bigl\{
A^\lambda_k \zeta^+_{-\vec k}(t_0) h_{\vec k}^{\lambda+}(t_0)
+ \cdots \bigr\} 
\nonumber \\
&\!\!\!\!\!\!=\!\!\!\!\!\!& 
- \int {d^3\vec k\over (2\pi)^3}\, \bigl\{
A^\lambda_k \zeta_k(t_0) h_k^*(t_0)\, 
a_{\vec k}^{\hbox{\tiny $(+)$}}
a_{\vec k}^{\lambda\hbox{\tiny $(+)$}\dagger}
+ \cdots \bigr\} .
%\nonumber 
\end{eqnarray}
For modes well within the horizon at the initial time, $|k\eta_0|\gg 1$, using the leading de Sitter forms for the mode functions, this becomes, 
\begin{equation}
S_0 = - \int {d^3\vec k\over (2\pi)^3}\, \biggl\{ A^\lambda_k 
{1\over\sqrt{4\epsilon}} {H^2\over M_{\rm pl}^2} {(k\eta_0)^2\over k^3} \, 
a_{\vec k}^{\hbox{\tiny $(+)$}}
a_{\vec k}^{\lambda\hbox{\tiny $(+)$}\dagger}
+ \cdots \biggr\} .
\end{equation}
This suggests that by defining dimensionless structure functions of the form
\begin{equation}
a_k^{+,\times} 
= {1\over\sqrt{4\epsilon}} {H^2\over M_{\rm pl}^2} k^2\eta_0^2 {A_k^{+,\times}\over k^3} 
\quad\hbox{and}\quad
b_k^{+,\times} 
= {1\over\sqrt{4\epsilon}} {H^2\over M_{\rm pl}^2} k^2\eta_0^2
{B_k^{+,\times}\over k^3} ,
\end{equation}
these functions describe how we have excited the different $\vec k$ modes in the initial state.  Of course, we are always free to define other coefficients however we wish through a dimensionless combination of the mass scales.  For example, we could have simply rescaled the initial state structure functions by $\zeta_k(t_0)$ and $h_k(t_0)$ or their complex conjugates; but unless the modes responsible for the structures that we observe in the CMB were just within the horizon at $t_0$, there will not be much of a difference between using the full mode functions and the limiting form that we have chosen.

Although we have been working so far with a general state, the fact that the Bunch-Davies vacuum gives a good explanation of the observed angular power spectra for the currently measured precisions suggests that we might wish to start by considering relatively small departures from the Bunch-Davies state.  Since we have already defined our initial state with respect to the vacuum, a small departure corresponds to choosing the dimensionless parameters to be small, $|a_k^{+,\times}|,|b_k^{+,\times}|\ll 1$.  Note that since the CMB is only sensitive to a relatively narrow range---in a logarithmic sense---of values of $k$, we could consider more interesting cases too, including those where $a_k^{+,\times}$ and $b_k^{+,\times}$ are increasing functions of $k$ over the range of momenta relevant for the CMB as long as $a_k^{+,\times},b_k^{+,\times}\to 0$ sufficiently rapidly for large values of $k$.

Once we have expanded the two-point functions to linear order in $a_k^{+,\times}$ and $b_k^{+,\times}$, it is convenient to make one last redefinition of them by combining them into a single pair of complex functions, 
\begin{equation}
c_k^+ \equiv a_k^+ +ib_k^+
\qquad\hbox{and}\qquad
c_k^\times \equiv a_k^\times +ib_k^\times , 
\end{equation}
when possible.  At this order in the initial state structure functions, the entangled power spectra are 
\begin{eqnarray}
{\langle \zeta_{-\vec k}(t)\zeta_{\vec k}(t)\rangle\over |\zeta_k(t)|^2} &=& 
1 + 2|c_k^+|^2 + 2|c_k^\times|^2 
+ C(k\eta_0) 
\Bigl[ (c^+_k)^2 + (c^{+*}_k)^2 + (c^\times_k)^2 + (c^{\times *}_k)^2 \Bigr]
 \\
&&
+ i S(k\eta_0)
\Bigl[ (c^+_k)^2 - (c^{+*}_k)^2 + (c^\times_k)^2 - (c^{\times *}_k)^2 \Bigr]
+ \cdots
\nonumber \\
{\langle h^+_{-\vec k}(t)h^+_{\vec k}(t)\rangle\over |h_k(t)|^2} &=& 
1 + 2 \bigl| c_k^+\bigr|^2 
+ C(k\eta_0) \Bigl[ (c_k^+)^2 + (c_k^{+*})^2\Bigr]
+ i S(k\eta_0) \Bigl[ (c_k^+)^2 - (c_k^{+*})^2 \Bigr]
+ \cdots 
\nonumber \\
{\langle h^\times_{-\vec k}(t)h^\times_{\vec k}(t)\rangle\over |h_k(t)|^2} &=& 
1 + 2 \bigl| c_k^\times \bigr|^2
+ C(k\eta_0) \Bigl[ (c_k^\times)^2 + (c_k^{\times *})^2\Bigr]
+ i S(k\eta_0) \Bigl[ (c_k^\times)^2 - (c_k^{\times *})^2 \Bigr]
+ \cdots 
\nonumber \\
{\langle h^+_{-\vec k}(t)h^\times_{\vec k}(t)\rangle\over |h_k(t)|^2} &=& 
c_k^+ c_k^{\times *} + c_k^\times c_k^{+*}
+ C(k\eta_0) \Bigl[ c_k^+ c_k^\times + c_k^{+*} c_k^{\times *} \Bigr]
+ i S(k\eta_0) \Bigl[ c_k^+ c_k^\times - c_k^{+*} c_k^{\times *} \Bigr]
+ \cdots
\nonumber \\
{\langle \zeta_{-\vec k}(t)h^+_{\vec k}(t)\rangle\over |\zeta_k(t)| |h_k(t)|} &=& 
- i\, C(k\eta_0) \bigl[ a_k^+ - a_k^{+*} \bigr]
+ S(k\eta_0) \bigl[ a_k^+ + a_k^{+*} \bigr]
+ 2b_k^+ 
\nonumber \\
&&
-\, i\, |c_k^+|^2 (c_k^+ - c_k^{+*}) 
- i\,\bigl[ c_k^+ c_k^{\times*} + c_k^{+*} c_k^\times \bigr] 
(c_k^\times - c_k^{\times*}) 
\nonumber \\
&& 
+\, i\, C(k\eta_0) \Bigl[ 
\bigl( (c_k^+)^2 + (c_k^{+*})^2 \bigr) (c_k^+ - c_k^{+*}) 
+ \bigl( c_k^+ c_k^\times + c_k^{+*} c_k^{\times*} \bigr) (c_k^\times - c_k^{\times*}) 
\Bigr]
\nonumber \\
&& 
-\, S(k\eta_0) \Bigl[ 
\bigl( (c_k^+)^2 - (c_k^{+*})^2 \bigr) (c_k^+ - c_k^{+*}) 
+ \bigl( c_k^+ c_k^\times - c_k^{+*} c_k^{\times*} \bigr) (c_k^\times - c_k^{\times*}) 
\Bigr]
+ \cdots
\nonumber
\end{eqnarray}
The functions $C(k\eta_0)$ and $S(k\eta_0)$ appearing in these expressions are 
\begin{eqnarray}
C(k\eta_0) &\equiv& {[(k\eta_0)^2-1] \cos 2k\eta_0 
- 2k\eta_0 \sin 2k\eta_0 \over (k\eta_0)^2+1}
\nonumber \\
S(k\eta_0) &\equiv& {[(k\eta_0)^2-1] \sin 2k\eta_0 
+ 2k\eta_0\cos 2k\eta_0\over (k\eta_0)^2+1} ; 
%\nonumber 
\end{eqnarray}
they are the oscillatory relics of the $e^{\pm ik\eta_0}$ factors in the mode functions.  In the limit where a mode starts well within the horizon, they become just cosines and sines; that is,  as $|k\eta_0|$ grows large,
\begin{equation}
C(k\eta_0)\to\cos 2k\eta_0
\qquad\hbox{and}\qquad 
S(k\eta_0)\to\sin 2k\eta_0 .
\end{equation}

We have also rescaled the two-point functions by dividing each by its standard Bunch-Davies form in the limit where we neglect the running of the spectral index, 
\begin{equation}
\lim_{t\to\infty}|\zeta_k(t)|^2 = {1\over 4\epsilon} {H^2\over M_{\rm pl}^2} {1\over k^3} + \cdots 
\qquad\hbox{and}\qquad 
\lim_{t\to\infty}|h_k(t)|^2 = {H^2\over M_{\rm pl}^2} {1\over k^3} + \cdots . 
\end{equation}
Such a rescaling allows us to see directly how the initial state has modified the primordial spectrum.

\section{Angular power spectra} 

\noindent 
While we cannot observe the power spectra of the primordial fluctuations directly, we are able to see their influences on the cosmic microwave background radiation and on the distribution of structures in the universe.  The primordial fluctuations essentially provide a set of initial conditions for the classical evolution of the universe which leads over time to distinct patterns in the fluctuations in the temperature ($T$) and polarisations ($E$ and $B$) of the CMB.  The $E$-polarised radiation represents the component that can be written as a divergence, whereas the $B$-polarised component is the divergenceless, `pure curl', part.  The physical principles that connect a set of initial conditions to a prediction for the correlations amongst the fluctuations in the CMB are well understood and moreover can be treated linearly---as a small spatially dependent perturbation to an otherwise smooth, isotropic background \cite{Ma:1995ey,Seljak:1996is,Zaldarriaga:1996xe,Hu:1997hp,Weinberg:2008zzc}.

Since we observe the cosmic microwave background radiation as it appears projected onto the sky, it is convenient to decompose the fluctuations in the CMB into spherical harmonics, 
\begin{equation}
a_{lm}^X = \int d\Omega\; X(\eta,\vec x;\theta,\varphi) Y_{lm}^*(\theta,\varphi;\hat e) .
\end{equation}
Here `$X$' refers to any of the $T$, $E$, or $B$ fluctuations in the CMB and $\hat e$ is the direction with respect to which we are defining the spherical harmonics.  The relation between the predicted form of a particular sort of fluctuation in the CMB and the primordial scalar or tensor fluctuations is given by the relation,
\begin{equation}
X(\eta,\vec x;\theta,\varphi) = \int {d^3\vec k\over (2\pi)^3}\; 
e^{i\vec k\cdot\vec x}\, \biggl\{ 
\sum_{\ell=0}^\infty \sum_{s=0,\pm 2} R^s_{\vec k} 
\Delta_{\ell s}^X(\eta,k) Y_{\ell s}(\theta,\varphi;\hat k)
\biggr\} .
\end{equation}
We have written the different fluctuations using a common symbol, $R^s_{\vec k}$, to include at once all of the primordial fluctuations that we are considering,
\begin{equation}
R^0_{\vec k} \equiv \zeta_{\vec k}(\eta_*)
\qquad\hbox{and}\qquad 
R^{\pm 2}_{\vec k} \equiv h_{\vec k}^{\pm 2}(\eta_*) .
\end{equation}
In writing these expressions we have also switched from the $+,\times$ polarisations of the tensor fluctuations to their $\pm 2$ helicity eigenstates.  These two bases for the tensor fluctuations are related by
\begin{equation}
h_{\vec k}^{+2}(\eta_*) = {\textstyle{1\over\sqrt{2}}} 
\bigl[ h_{\vec k}^+(\eta_*) - i h_{\vec k}^\times(\eta_*) \bigr] 
\qquad\hbox{and}\qquad
h_{\vec k}^{-2}(\eta_*) 
= {\textstyle{1\over\sqrt{2}}} 
\bigl[ h_{\vec k}^+(\eta_*) + i h_{\vec k}^\times(\eta_*) \bigr] . 
\end{equation}

The other set of functions that appear in the above expression, $\Delta_{\ell s}^X(\eta,k)$, are the standard {\it transfer functions\/}\footnote{See chapter 6 of \cite{Weinberg:2008zzc} or chapter 7 of \cite{Dodelson:2003ft}, for example.}; they tell how the primordial fluctuations produced at some extremely early time $\eta_*$, such as the end of the inflationary expansion, evolve and contribute to the `$X$' fluctuations in the CMB at $\eta$.  These transfer functions are calculated using a linearised form of Einstein's equations plus the Boltzmann equations that describe how the different ingredients of the universe interact with each other.  One useful property, which we can state here without going into a detailed treatment of them, is that the transfer functions satisfy basic parity relations.  The $T$ fluctuations and the $E$ component of the polarisation, for example, are both even under $s\to -s$, while the $B$ component is odd under $s\to -s$, 
\begin{equation}
\Delta_{\ell,-s}^T(\eta,k) = \Delta_{\ell s}^T(\eta,k) , 
\qquad 
\Delta_{\ell,-s}^E(\eta,k) = \Delta_{\ell s}^E(\eta,k) , 
\qquad 
\Delta_{\ell,-s}^B(\eta,k) = -\Delta_{\ell s}^B(\eta,k) . 
\end{equation}
The last relation holds because when we reverse the spin, curls circulate in the opposite directions.

The spherical harmonics that appear in the expansions for $a_{lm}^X$ and $X(\eta,\vec x;\theta,\varphi)$ are defined relative to different reference axes.  We are free to express one sort of spherical harmonics in terms of the other through a suitable `change of basis' formula,
\begin{equation}
Y_{\ell m}(\theta,\varphi;\hat k) = \sqrt{{4\pi\over 2l+1}} \sum_{m'} 
{}_{-m}Y^*_{\ell m'}(\hat k;\hat e) Y_{\ell m'}(\theta,\varphi;\hat e) .
\end{equation}
The functions ${}_{-m}Y^*_{\ell m'}(\hat k;\hat e)$ that relate them are the {\it spin-weighted spherical harmonics\/}.  Since there are only spin 0 and spin 2 fluctuations, we shall only be requiring ${}_{0}Y_{lm}(\hat k;\hat e)$ and ${}_{\pm 2}Y_{lm}(\hat k;\hat e)$.  The former correspond to the ordinary spherical harmonics, ${}_{0}Y_{lm}(\hat k;\hat e)=Y_{lm}(\hat k;\hat e)$, but the latter are more complicated.  However, ${}_{\pm 2}Y_{lm}(\theta,\varphi)$ can be expressed in terms of the familiar spherical harmonics through the relation ${}_{\pm 2}Y_{lm}(\theta,\varphi) = (\hat A\pm i\hat B)Y_{lm}(\theta,\varphi)$ where
\begin{equation}
\hat A = {-\cos\theta\sin\theta\, \partial_\theta + \sin^2\theta\, \partial_\theta^2 
- \partial_\varphi^2\over\sin^2\theta\sqrt{(l-1)l(l+1)(l+2)}},
\qquad
\hat B = {2\sin\theta\, \partial_\theta\partial_\varphi 
- 2\cos\theta\, \partial_\varphi\over\sin^2\theta\sqrt{(l-1)l(l+1)(l+2)}} , 
\end{equation}
which can also be expressed directly as a linear combination of the ordinary spherical harmonics, 
\begin{eqnarray}
&&\hskip-0.3125truein
{}_{\pm 2}Y_{lm}(\theta,\varphi) 
\nonumber \\
&=& 
{1\over\sin^2\theta} \biggl\{ 
{1\over 2l+3} \biggl[ {(l-1)l(l+m+1)(l-m+1)(l+m+2)(l-m+2)\over 
(l+1)(l+2)(2l+1)(2l+5)} \biggr]^{1/2}
Y_{l+2,m}(\theta,\varphi) 
\nonumber \\
&&\qquad\quad
\mp 2m \biggl[ {(l-1)(l+m+1)(l-m+1)\over l(l+1)(l+2)(2l+1)(2l+3)} \biggr]^{1/2}
Y_{l+1,m}(\theta,\varphi) 
\nonumber \\
&&\qquad\quad
+ {2 \bigl[ 3m^2 -l(l+1) \bigr]\over (2l-1)(2l+3)} 
\biggl[ {(l-1)(l+2)\over l(l+1)} \biggr]^{1/2}Y_{lm}(\theta,\varphi) 
\nonumber \\
&&\qquad\quad
\pm 2m \biggl[ {(l+2)(l+m)(l-m)\over (l-1)l(l+1)(2l+1)(2l-1)} \biggr]^{1/2}
Y_{l-1,m}(\theta,\varphi) 
\nonumber \\
&&\qquad\quad
+ {1\over 2l-1} \biggl[ {(l+1)(l+2)(l+m)(l-m)(l+m-1)(l-m-1)\over 
(l-1)l(2l-3)(2l+1)} \biggr]^{1/2}
Y_{l-2,m}(\theta,\varphi) \biggr\} .
\nonumber  
\end{eqnarray}
Notice that when $m=0$, ${}_{-2}Y_{l0}(\theta,\varphi)={}_2Y_{l0}(\theta,\varphi)$.  In the following, it is convenient further to absorb the normalisation factor appearing above by rescaling the transfer functions,
\begin{equation}
\tilde\Delta_{\ell s}^X(\eta,k) \equiv \sqrt{{4\pi\over 2\ell+1}} \Delta_{\ell s}^X(\eta,k) . 
\end{equation}
Substituting the expression for $X$ in terms of the transfer functions into $a_{lm}^X$ and then integrating over the angles, yields the more practical form,
\begin{equation}
a_{lm}^X = \int {d^3\vec k\over (2\pi)^3}\; e^{i\vec k\cdot\vec x}\, \biggl\{ 
\sum_{s=0,\pm 2} R^s_{\vec k} \tilde\Delta_{ls}^X(\eta,k) \biggr\} 
{}_{-s}Y^*_{lm}(\hat k;\hat e) .
\end{equation}

There remain only the angular two-point functions to evaluate between different $a_{lm}^X$'s, 
\begin{equation}
C^{X,X'}_{ll',mm'} = \langle a_{lm}^X a_{l'm'}^{X'*}\rangle . 
\end{equation}
This can be done once we have provided the corresponding correlators for the primordial power spectra, 
\begin{equation}
\langle R^s_{\vec k} R^{s'*}_{\vec k'} \rangle 
= {2\pi^2\over k^3}\, P_k^{ss'}\, (2\pi)^3\, \delta^3(\vec k-\vec k') . 
\end{equation}
Up to the change in the polarisation basis that we made for the spin 2 fluctuations, these power spectra correspond to those derived in the previous section.

Most generally, we could imagine a primordial power spectrum that depends on the direction of $\vec k$ in addition to the mixing introduced here; but for simplicity, we have been assuming that the dependence of $P_k^{ss'}$ is only on the magnitude of $\vec k$.  The spin-weighted spherical harmonics satisfy the following orthogonality relation when their `spin' indices are equal, 
\begin{equation}
\int d\Omega_{\hat k}\; \Bigl\{ {}_{-s}Y^*_{lm}(\hat k;\hat e) 
{}_{-s}Y_{l'm'}(\hat k;\hat e) \Bigr\} 
= \delta_{ll'}\delta_{mm'} . 
\end{equation}
For this reason it is instructive to separate the diagonal ($s=s'$) from the off-diagonal ($s\not=s'$) cases when writing the predictions for $C^{X,X'}_{ll',mm'}$,
\begin{eqnarray}
C^{X,X'}_{ll',mm'} &=& 
{1\over 4\pi} \delta_{ll'}\delta_{mm'} \int_0^\infty {dk\over k}\, 
\sum_{s} \Bigl\{ 
P_k^{ss} \tilde\Delta_{ls}^X(\eta,k) \tilde\Delta_{ls}^{X'}(\eta,k) \Bigr\}  
 \\
&& 
+ {1\over 4\pi} \int_0^\infty {dk\over k}\, 
\sum_{s\not=s'} \Bigl\{ 
P_k^{ss'}\tilde\Delta_{ls}^X(\eta,k) \tilde\Delta_{l' s'}^{X'}(\eta,k) \Bigr\} 
\int d\Omega_{\hat k}\; 
\Bigl\{ {}_{-s}Y^*_{lm}(\hat k;\hat e) 
{}_{-s'}Y_{l'm'}(\hat k;\hat e) \Bigr\} , 
\nonumber 
\end{eqnarray}
where $s$ and $s'$ assume the values $0$, $2$, or $-2$.  Note that since ${}_sY_{lm}(\theta,\varphi)\propto e^{im\varphi}$, the correlation functions are still diagonal in $m$ and $m'$, 
\begin{equation}
\int d\Omega_{\hat k}\; 
\Bigl\{ {}_{-s}Y^*_{lm}(\hat k;\hat e) {}_{-s'}Y_{l'm'}(\hat k;\hat e) \Bigr\} 
= \delta_{mm'}\, I^{ll',m}_{ss'} .
\end{equation}
We also have trivially that $(I^{ll',m}_{ss'})^*=I^{l'l,m}_{s's}$.

In the standard inflationary picture, the primordial power spectra do not mix different types of fluctuations, $P_k^{ss'}=0$ for $s\not=s'$.  This in turn means that the angular power spectra are diagonal in their angular indices, $C^{X,X'}_{ll',mm'} \propto \delta_{ll'}\delta_{mm'}$.  When there is some entanglement between the scalar and tensor fluctuations in the initial state, the angular power spectra no longer need be diagonal in this sense.  In this case there are in fact {\it two\/} ways in which the two-point correlators can depart from the usual scenario:  (i) in the $k$-dependence of the power spectra, even the diagonal ones such as $P_k^{00}$ are affected by the structure of the initial state, and (ii) in non-vanishing correlations between different values of $l$ and $l'$ in the angular power spectra.  Note that a nonstandard $k$-dependence---in the $TT$ power spectrum, for instance---can just as easily be generated---or perhaps even more easily generated---by modifying the state of the $\zeta(t,\vec x)$ field alone, but that the latter signature is particularly distinctive of having some broken rotational invariance.

\subsection{$TT$ angular correlations} 

\noindent
As an example, let us consider the $TT$ correlation function, 
\begin{eqnarray}
C^{TT}_{ll',mm'} &=& 
{1\over 4\pi} \delta_{ll'}\delta_{mm'} \int_0^\infty {dk\over k}\, 
\Bigl\{ 
P_k^{00} \tilde\Delta_{l0}^T(\eta,k) \tilde\Delta_{l0}^T(\eta,k) 
+ \bigl[ P_k^{22} + P_k^{-2,-2} \bigr] \tilde\Delta_{l2}^T(\eta,k) \tilde\Delta_{l2}^T(\eta,k) 
\Bigr\}  
\nonumber \\
&& 
+ {1\over 4\pi} \delta_{mm'} \int_0^\infty {dk\over k}\, 
\Bigl\{ 
\bigl[ P_k^{02}I^{ll',m}_{02} + P_k^{0,-2}I^{ll',m}_{0,-2} \bigr] 
\tilde\Delta_{l0}^T(\eta,k) \tilde\Delta_{l'2}^T(\eta,k) \Bigr\}   
\nonumber \\
&& 
+ {1\over 4\pi} \delta_{mm'} \int_0^\infty {dk\over k}\, 
\Bigl\{ 
\bigl[ P_k^{20}I^{ll',m}_{20} + P_k^{-20}I^{ll',m}_{-20} \bigr] 
\tilde\Delta_{l2}^T(\eta,k) \tilde\Delta_{l'0}^T(\eta,k) \Bigr\}   
\nonumber \\
&& 
+ {1\over 4\pi} \delta_{mm'} \int_0^\infty {dk\over k}\, 
\Bigl\{ 
\bigl[ P_k^{2,-2}I^{ll',m}_{2,-2} + P_k^{-2,2}I^{ll',m}_{-2,2} \bigr] 
\tilde\Delta_{l2}^T(\eta,k) \tilde\Delta_{l'2}^T(\eta,k) \Bigr\} . 
%\nonumber 
\end{eqnarray}
In the standard, statistically isotropic models where $P^{ss'}\propto\delta^{ss'}$, only the first line of this expression contributes and the correlations are diagonal in both $ll'$ and $mm'$.  But when there is some generic scalar-tensor entanglement, there are additional contributions.  The largest effect in the diagonal entries, $l=l'$, still comes from the $P_k^{00}$ contribution, though even the diagonal correlators will have additional contributions because the spin-weighted spherical harmonics are generally not orthogonal to the ordinary spherical harmonics.  For example,\footnote{\cite{Samaddar} contains some useful integrals (and their derivations) of associated Legendre functions.} when $m\not=0$,
\begin{eqnarray}
I^{ll,m}_{\mp20} 
&=& 
\int d^2\Omega\, \bigl({}_{\pm 2}Y_{lm}(\theta,\phi)\bigr)^* 
Y_{lm}(\theta,\phi) 
\nonumber \\ 
&=& 
{1\over 2m} \biggl[ {(l-1)(l+2)\over l(l+1)} \biggr]^{1/2} 
\biggl\{ {(l-m+1)(l-m+2)\over 2l+3} {l\over (l+2)} 
+ {2 \bigl[ 3m^2 -l(l+1) \bigr](2l+1)\over (2l-1)(2l+3)} 
\nonumber \\
&&\qquad\qquad\qquad\qquad\qquad\quad
+ {(l-m)(l-m-1)\over 2l-1} {(l+1)\over (l-1)} 
 \biggr\} ;
%\nonumber 
\end{eqnarray}
in particular, the $m=l$ and $m=l-1$ cases are 
\begin{eqnarray}
\int d^2\Omega\, \bigl({}_{\pm 2}Y_{ll}(\theta,\phi)\bigr)^* Y_{ll}(\theta,\phi)
&=& 
\sqrt{{(l-1)(l+1)\over l(l+2)}}
\nonumber \\ 
\int d^2\Omega\, \bigl({}_{\pm 2}Y_{l,l-1}(\theta,\phi)\bigr)^* Y_{l,l-1}(\theta,\phi) 
&=& 
{l^2-2l-2\over\sqrt{(l-1)l(l+1)(l+2)}} .
%\nonumber 
\end{eqnarray}
And when $m=0$,
\begin{equation}
\int d^2\Omega\, \bigl({}_{\pm 2}Y_{l0}(\theta,\phi)\bigr)^* Y_{l0}(\theta,\phi) 
= - \sqrt{{l(l-1)\over(l+1)(l+2)}} .
\end{equation}

As an illustration, consider the case when $m=m'=0$.  Looking at the diagonal entries of the $TT$ correlation function, we have 
\begin{eqnarray}
C^{TT}_{ll,00} &=& 
{1\over 4\pi} \int_0^\infty {dk\over k}\, 
\Bigl\{ 
P_k^{00} \tilde\Delta_{l0}^T(\eta,k) \tilde\Delta_{l0}^T(\eta,k) 
+ 2P_k^{++} \tilde\Delta_{l2}^T(\eta,k) \tilde\Delta_{l2}^T(\eta,k) 
\Bigr\}  
\nonumber \\
&& 
- {1\over 2\pi} \sqrt{{2l(l-1)\over(l+1)(l+2)}}\int_0^\infty {dk\over k}\, 
P_k^{0+} \tilde\Delta_{l0}^T(\eta,k) \tilde\Delta_{l2}^T(\eta,k) .
%\nonumber 
\end{eqnarray}
So the terms that contribute to these entries of the angular correlation functions have different structures from those that would have appeared in an isotropic model.

However, the most distinctive case occurs when $l\not=l'$.  Since the overlaps between the spin-weighted spherical harmonics do not vanish generically, $C^{TT}_{ll',mm'}$ could contain nonvanishing off-diagonal elements as well.  As an example, when $l'=l+1$ and $m\not=0$, the overlap integral between the $s=\pm 2$ and $s'=0$ spin-weighted spherical harmonics is 
\begin{equation}
\int d^2\Omega\, \bigl({}_{\pm 2}Y_{lm}(\theta,\phi)\bigr)^* 
Y_{l+1,m}(\theta,\phi) 
= \mp (m-1)\biggl[ {(2l+1)(2l+3)(l-m+1)\over (l-1)l(l+1)(l+2)(l+m+1)} \biggr]^{1/2} . 
\end{equation}
\vskip12truept

\subsection{Oscillations} 

\noindent
A very general initial state leads similarly to a very general set of angular correlation functions for the CMB.  As we mentioned from the outset, the most constructive approach would be to use observations to constrain the scale dependence of the primordial power spectrum.  This would in turn constrain the possible $k$-dependence of the entanglement functions, $A_k^\lambda$ and $B_k^\lambda$.  But since doing so would be a little too ambitious even for the precision of the current measurements of the CMB, we shall instead plot the leading behaviour of a `typical' case within our class of states where the initial state contains only a tiny amount of entanglement between the scalar and tensor fluctuations.

Suppose that we have expanded to leading order in the structures of the initial action; the primordial power spectra will then have the forms given at the end of section 3.  In principle, the entanglement in this case is parametrised by six real functions, the real and imaginary parts of $a_k^+$ and $a_k^\times$, $b_k^+$ and $b_k^\times$, and the initial time $\eta_0$.  However, a small change in the initial time $\eta_0\to\eta_0+\delta\eta_0$ can be absorbed by redefining the other six functions.  To simplify the picture further, let us assume that these six functions are approximately constant over the range of $k$'s relevant for cosmological observations.  This still leaves six parameters to fix; but a {\it generic\/} choice leads to a similar qualitative picture for the new features in the angular correlations of the CMB.

Because the entanglement couples the scalar and tensor fluctuations, unless there is some fine-tuning in the structure functions, $A_k^\lambda$ and $B_k^\lambda$, the effect of the entanglement on each of the power spectra $P^{ss'}$ is similar.  As an example, in the limit where the rescaled entanglement parameters are tiny, the scalar-scalar and the symmetric part of the tensor-tensor power spectra receive exactly the same corrections,
\begin{eqnarray}
P_k^{00} &=& 
{\cal A}_s \biggl( {k\over k_*} \biggr)^{n_s-1} \Bigl\{ 
1 + 2|c_k^+|^2 + 2|c_k^\times|^2 
+ C(k/k_0) \Bigl[ (c_k^+)^2 + (c_k^{+*})^2 + (c_k^\times)^2 + (c_k^{\times *})^2 \Bigr]
\nonumber \\
&&\qquad\qquad\quad
- i S(k/k_0) \Bigl[ (c_k^+)^2 - (c_k^{+*})^2 + (c_k^\times)^2 - (c_k^{\times *})^2 \Bigr]
+ \cdots \Bigr\}
\end{eqnarray}
and
\begin{eqnarray}
&&\hskip-0.5truein
P_k^{22} + P_k^{-2,-2} = 
P_k^{++} + P_k^{\times\times}
\nonumber \\
&=& 
{1\over 2} {\cal A}_t \biggl( {k\over k_*} \biggr)^{n_t} \Bigl\{ 
2 + 2|c_k^+|^2 + 2|c_k^\times|^2 
+ C(k/k_0) \Bigl[ (c_k^+)^2 + (c_k^{+*})^2 + (c_k^\times)^2 + (c_k^{\times *})^2 \Bigr]
\nonumber \\
&&\qquad\qquad\quad
- i S(k/k_0) \Bigl[ (c_k^+)^2 - (c_k^{+*})^2 + (c_k^\times)^2 - (c_k^{\times *})^2 \Bigr]
+ \cdots \Bigr\} .
%\nonumber
\end{eqnarray}
In writing these expressions, and in making their associated plots, we have introduced some of the standard parameters used to fit the fluctuations in the CMB.  ${\cal A}_s$ and ${\cal A}_t$ are the amplitudes for the pure Bunch-Davies parts of the fluctuations and $n_s$ and $n_t$ are the indices associated with the running of the power spectra. $k_*$ is a pivot scale chosen when performing fits to CMB observations.  Also, for the purpose of plotting the effect of the entanglement on the CMB, it is useful to recast the initial time as a wave number, $\eta_0^{-1}=-k_0$.  Notice that on average, the new terms tend to increase the amplitude; so to match the usual fits without the entanglement we ought to decrease the amplitudes ${\cal A}_s$ and ${\cal A}_t$ slightly to compensate, e.g.
\begin{equation}
{\cal A}_s \to {{\cal A}_s\over 1 + 2|c_k^+|^2 + 2|c_k^\times|^2} . 
\end{equation}

\begin{figure}
\includegraphics[width=4.5truein]{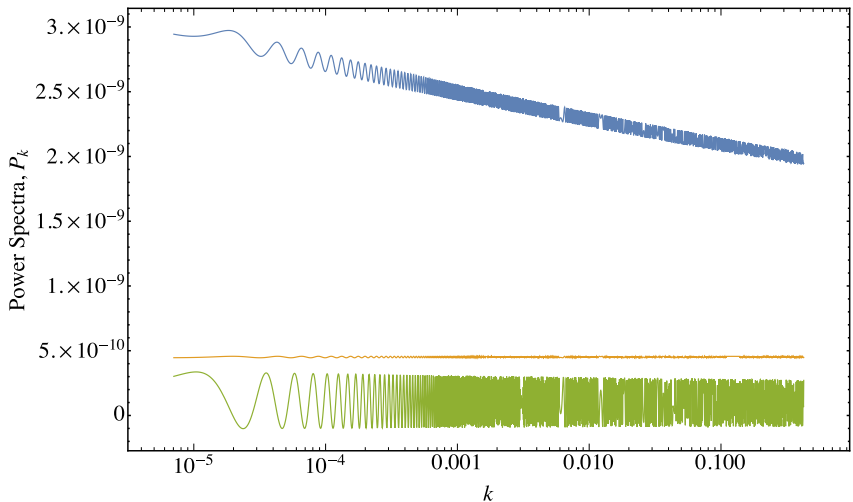}
\caption{Three basic power spectra.  The primordial power spectrum for the scalar fluctuations, $P_k^{00}$, is shown in blue, for the tensors, $P_k^{++}$, in yellow, and for the mixed scalar-tensor power spectrum, $P_k^{0+}$, in light green.  In making these plots, we have chosen the fairly standard values ${\cal A}_s = 2.285\cdot 10^{-9}$, ${\cal A}_t = r\, {\cal A}_s$ (where $r=0.2$ is the tensor-to-scalar ratio), $n_s=0.9619123$, and $n_t=0$.  The pivot scale is $k_*=0.01$.  Finally, the parameters chosen to describe the initial entanglement between the scalar and tensor fluctuations are $c_k^+=0.08$, $c_k^\times=0.08$, $b_k^+=0.05$, and $\eta_0^{-1} = - 7.052713\cdot 10^{-6}$.}
\end{figure}

\begin{figure}
\includegraphics[width=4.5truein]{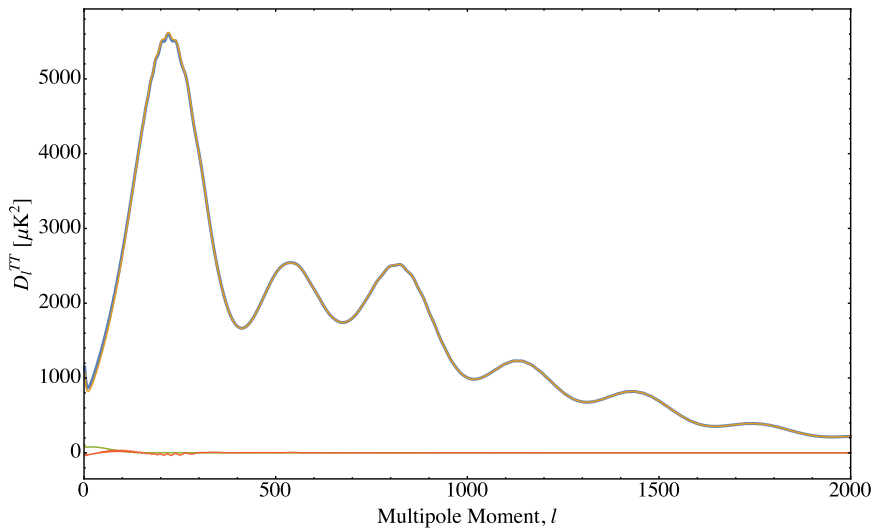}
\caption{The angular $TT$ correlations that result from the choice of parameters described in figure 1.  Note that what is plotted here is based on $C_{ll,00}^{TT}$ multiplied by $l(l+1)/2\pi$ and {\it not\/} summed over $m$, $D_l^{TT}\sim {l(l+1)\over 2\pi}C_{ll,00}^{TT}$.  This figure was generated using the {\sc class\/} numerical package \cite{Lesgourgues:2011re,Blas:2011rf}.  The blue curve (nearly covered up by the yellow curve) shows the total result, while the yellow, light green, and red curves show the contributions from the $P_k^{00}$, $P_k^{++}$, and $P_k^{0+}$ primordial fluctuations respectively. }
\end{figure}

In figures 1--4 we illustrate the effect of the entanglement for a particular choice of the parameters, $a_k^{+,\times}$, $b_k^{+,\times}$, which we have chosen to be all constant for simplicity.  The effect of the entanglement is slightly larger in the scalar-tensor cross-correlation since it is {\it linear\/} in the parameters, although its {\it overall\/} contribution is suppressed in the perturbative limit since it does not have a zeroth order, Bunch-Davies, contribution.  An example of a set of primordial power spectra are shown in figure 1.  Figure 2 shows the angular $TT$ correlations generated by these primordial power spectra.  To produce these figures we used the transfer functions computed using the {\sc class\/} numerical routines \cite{Lesgourgues:2011re,Blas:2011rf}.  Since the overlap integrals between different spin-weighted spherical harmonics are difficult to evaluate analytically, we have illustrated the case when $l=l'$ and $m=m'=0$ and then multiplied the resulting $C_{ll,00}^{TT}$ by $l(l+1)/2\pi)$ to be able to compare with the standard, nonentangled result.  In order to make clearer the role of the entanglement, in figure 3 we show the entangled result with the standard, statistically isotropic part subtracted from it.  To highlight the effect on the lowest multipole moments, we show this difference again in figure 4 for the lowest $l$'s.  For the particular choice of entanglement parameters that we made, there is generally a suppression in the correlation power up to about $l\sim 40$.  This could be made larger by increasing the amplitude of the entanglement parameters.

\begin{figure}
\includegraphics[width=4.5truein]{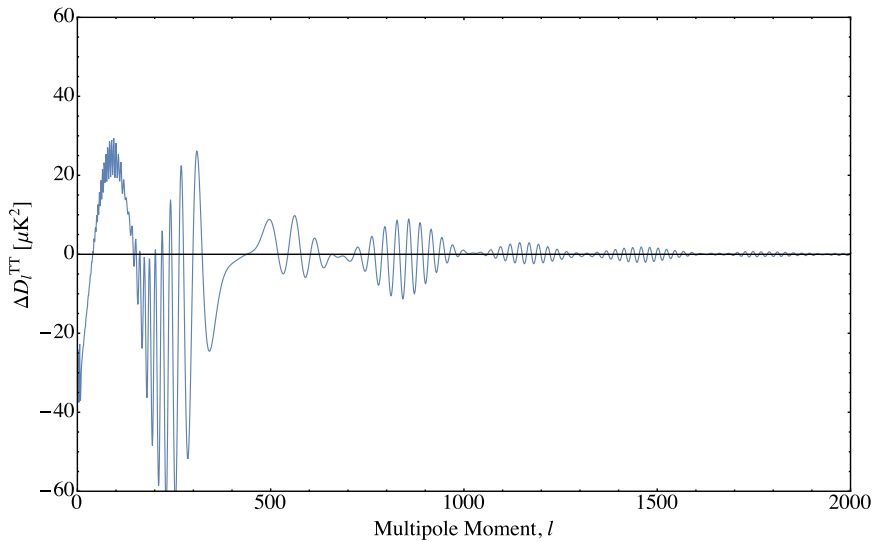}
\caption{Here we have plotted the difference between angular $TT$ correlations for the entangled state shown in the last figure and the standard unentangled case.  Since the entangled contribution oscillates about the Bunch-Davies power spectrum, the difference is largest at the peaks, as shown.}
\end{figure}

\begin{figure}
\includegraphics[width=4.5truein]{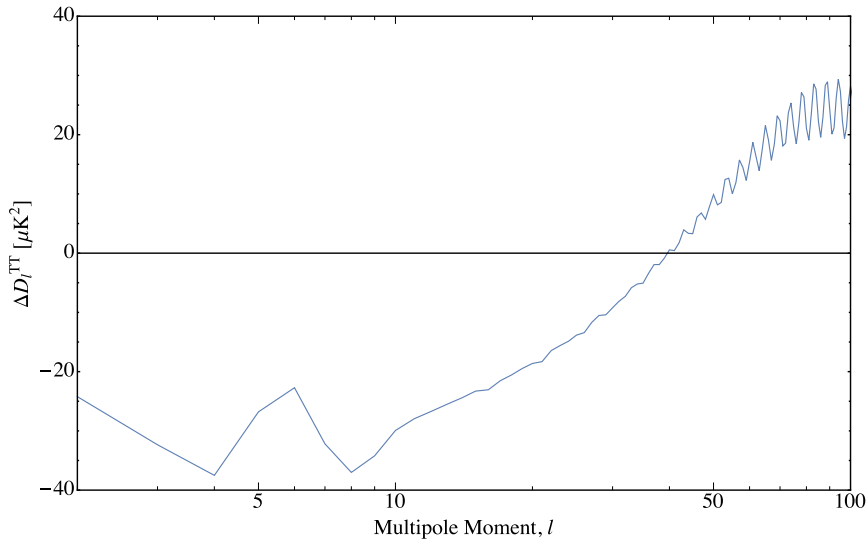}
\caption{Here we show the effect of the entanglement on the lowest multipole moments.  For this choice of the entanglement parameters, there is a slight suppression of power up to about $l\sim 40$.}
\end{figure}

Note that this is only one possibility for the angular correlations in this picture.  If instead of having chosen constant values for the entanglement function we had given them a scale dependence, we could enhance or diminish the angular power spectrum for either the large or the small $l$'s.

\section{Concluding remarks} 

\noindent 
In this article we have explored a few of the implications for a particular type of entanglement between the scalar and tensor fluctuations during inflation.  Since this entanglement has only been included within the state, the symmetry-breaking is comparatively mild---the underlying dynamics as described by the $3+1$ dimensional action is still rotationally invariant.  We showed how the propagators could be solved exactly when the initial state includes some simple bilinear structures  in the two types of fluctuations, which can be characterised by a set of functions:  two complex, $A_k^{+,\times}$, and two real, $B_k^{+,\times}$.  These propagators in turn led to exact expressions for the power spectra of the primordial fluctuations, which include cross-correlations between the scalar and tensors and between different polarisations of the tensor fluctuations.

Because the formalism allows in principle for very general possible structures in the state, it is difficult to make absolute statements about the possible observational signatures from such entanglements.  But when the state is close to the Bunch-Davies state it is reasonable to expect (i) that there will be non-vanishing off-diagonal angular correlations in the CMB, and (ii) that there can be rapid, synchronised oscillations in all of the components of the $C^{TT}_{ll',mm'}$ when the state is defined at an initial time $\eta_0$.  The frequency of these oscillations is related to $\eta_0$.  To keep the picture causal $|k\eta_0|$ should not be smaller than 1 for the wavenumbers relevant for the CMB.  For this reason it is not possible to tune the oscillations to suppress the power consistently over the lower $l$ multipole moments, since the apparent coherence in the observations is over a larger number of $l$'s than such a causal-limit would allow.  Of course, a suitable choice of the structures $a_k^{+,\times}$ and $b_k^{+,\times}$ themselves could be made to mimic the observations more closely.

It is important also to note that the oscillations that we have found are not a necessary or an inevitable consequence of the entanglement.  Other works {\cite{Kanno} have investigated other entangled theories in de Sitter backgrounds without producing oscillations.  Something similar to these cases could have been done in our picture by choosing the structure functions to cancel the $k\eta_0$ dependence completely from the mode functions that appear in the initial action.  This could be a natural way to define the state when the initial time is not meant to be physically important or when some other principle is being used to define the state than what was done here.

One of our goals in this article was to solve for the effect of a quadratic entanglement in the initial state on the primordial two-point functions {\it exactly\/}.  If we had wished instead to examine only states that are perturbatively close to the Bunch-Davies state, then we could also have considered cases with more severely broken symmetries.  For example, we have examined states whose structure functions still only depended on $k=|\!|\vec k|\!|$; more generally, we could also analyse states that depend on the direction of $\vec k$ as well.  This would lead to anisotropic power spectra, $P_{\vec k}^{ss'}$, which would be more directly akin to some of the anisotropic models that have been proposed.

We have described the full expression for the propagator in detail and set up an interaction picture for the time-evolution since our formalism could be used to study the possible structures for the state more fully and thereby extract a more complete picture of the universe during the inflationary epoch.  To do so, we should include an additional set of separate quadratic structures for each of the scalar and tensor fluctuations.  Having such an initial state would allow more natural fits to the CMB angular correlation functions, since these structures would appear already at {\it linear\/} order in the scalar-scalar primordial power spectrum, rather than at quadratic order, as we have seen for the purely entangled state.  A complete interaction picture would allow other scenarios to be treated as well.  A picture that included cubic structures in the initial state as well as cubic operators in the dynamical part of the action could be used to generate non-Gaussian structures in the CMB.  By combining all of these structures and working backwards, constraining them through cosmological observations, we should obtain a deeper picture of our universe and should be better able to deduce what principles and symmetries are needed to understand its earliest moments.

\acknowledgments

Tereza Vardanyan is grateful for the support of the Department of Energy (DE-FG03-91-ER40682).  We should both like to thank Rich Holman and Nadia Bolis for valuable discussions based on their similar work \cite{Bolis:2016vas}, which builds upon the states developed in \cite{Albrecht:2014aga}.  We are especially grateful to the authors of \cite{Chen:2014eua} for providing some of their Mathematica code which was essential for generating the figures in this article.


\vskip36truept


\begin{thebibliography}{99}

%\cite{Ade:2015lrj}
\bibitem{Ade:2015lrj}
P.~A.~R.~Ade {\it et al.} [Planck Collaboration], ``Planck 2015 results. XX. Constraints on inflation,'' arXiv:1502.02114 [astro-ph.CO].

%\cite{Ade:2015ava}
\bibitem{Ade:2015ava}
P.~A.~R.~Ade {\it et al.} [Planck Collaboration], ``Planck 2015 results. XVII. Constraints on primordial non-Gaussianity,'' arXiv:1502.01592 [astro-ph.CO].

%\cite{Agarwal:2012mq}
\bibitem{Agarwal:2012mq}
N.~Agarwal, R.~Holman, A.~J.~Tolley, and J.~Lin, ``Effective field theory and non-Gaussianity from general inflationary states,'' JHEP {\bf 1305} (2013) 085 [arXiv:1212.1172 [hep-th]].

%\cite{Collins:2013kqa}
\bibitem{Collins:2013kqa}
H.~Collins, ``Initial state propagators,'' JHEP {\bf 1311} (2013) 077 [arXiv:1309.2656 [hep-th]].

%\cite{Ade:2015hxq}
\bibitem{Ade:2015hxq}
P.~A.~R.~Ade {\it et al.} [Planck Collaboration], ``Planck 2015 results. XVI. Isotropy and statistics of the CMB,'' arXiv:1506.07135 [astro-ph.CO].

%\cite{Schwarz:2015cma}
\bibitem{Schwarz:2015cma}
D.~J.~Schwarz, C.~J.~Copi, D.~Huterer, and G.~D.~Starkman, ``CMB Anomalies after Planck,'' arXiv:1510.07929 [astro-ph.CO].

%\cite{Soda:2012zm}
\bibitem{Soda:2012zm}
Some review of anisotropic inflationary models can be found in \\
J.~Soda, ``Statistical Anisotropy from Anisotropic Inflation,'' Class.\ Quant.\ Grav.\  {\bf 29} (2012) 083001 [arXiv:1201.6434 [hep-th]];\\
%\cite{Maleknejad:2012fw}
%\bibitem{Maleknejad:2012fw}
A.~Maleknejad, M.~M.~Sheikh-Jabbari and J.~Soda, ``Gauge Fields and Inflation,'' Phys.\ Rept.\  {\bf 528} (2013) 161 [arXiv:1212.2921 [hep-th]].

%\cite{Watanabe:2010bu}
\bibitem{Watanabe:2010bu}
M.~a.~Watanabe, S.~Kanno, and J.~Soda, ``Imprints of Anisotropic Inflation on the Cosmic Microwave Background,'' Mon.\ Not.\ Roy.\ Astron.\ Soc.\  {\bf 412} (2011) L83 [arXiv:1011.3604 [astro-ph.CO]].
  
%\cite{Chen:2014eua}
\bibitem{Chen:2014eua}
X.~Chen, R.~Emami, H.~Firouzjahi, and Y.~Wang, ``The TT, TB, EB and BB correlations in anisotropic inflation,'' JCAP {\bf 1408} (2014) 027 [arXiv:1404.4083 [astro-ph.CO]].

%\cite{Cheung:2007st}
\bibitem{Cheung:2007st}
C.~Cheung, P.~Creminelli, A.~L.~Fitzpatrick, J.~Kaplan and L.~Senatore, ``The Effective Field Theory of Inflation,'' JHEP {\bf 0803} (2008) 014 [arXiv:0709.0293 [hep-th]].

%\cite{Weinberg:2008hq}
\bibitem{Weinberg:2008hq}
S.~Weinberg, ``Effective Field Theory for Inflation,'' Phys.\ Rev.\ D {\bf 77} (2008) 123541 [arXiv:0804.4291 [hep-th]].

%\cite{Bartolo:2015qvr}
\bibitem{Bartolo:2015qvr}
N.~Bartolo, D.~Cannone, A.~Ricciardone, and G.~Tasinato, ``Distinctive Signatures of Space-Time Diffeomorphism Breaking in EFT of Inflation,'' arXiv:1511.07414 [astro-ph.CO].

%\cite{Coleman:1998ti}
\bibitem{Coleman:1998ti}
S.~R.~Coleman and S.~L.~Glashow, ``High-energy tests of Lorentz invariance,'' Phys.\ Rev.\ D {\bf 59} (1999) 116008 [hep-ph/9812418].

%\cite{Weinberg:2003sw}
\bibitem{Weinberg:2003sw}
S.~Weinberg, ``Adiabatic modes in cosmology,'' Phys.\ Rev.\ D {\bf 67} (2003) 123504 [astro-ph/0302326].

%\cite{Maldacena:2002vr}
\bibitem{Maldacena:2002vr}
J.~M.~Maldacena, ``Non-Gaussian features of primordial fluctuations in single field inflationary models,'' JHEP {\bf 0305} (2003) 013 [astro-ph/0210603].

%\cite{Collins:2011mz}
\bibitem{Collins:2011mz}
H.~Collins, ``Primordial non-Gaussianities from inflation,'' arXiv:1101.1308 [astro-ph.CO].

%\cite{Weinberg:2005vy}
\bibitem{Weinberg:2005vy}
S.~Weinberg, ``Quantum contributions to cosmological correlations,'' Phys.\ Rev.\ D {\bf 72} (2005) 043514 [hep-th/0506236].

%\cite{Ma:1995ey}
\bibitem{Ma:1995ey}
C.~P.~Ma and E.~Bertschinger, ``Cosmological perturbation theory in the synchronous and conformal Newtonian gauges,'' Astrophys.\ J.\  {\bf 455} (1995) 7 [astro-ph/9506072].

%\cite{Seljak:1996is}
\bibitem{Seljak:1996is}
U.~Seljak and M.~Zaldarriaga, ``A Line of sight integration approach to cosmic microwave background anisotropies,'' Astrophys.\ J.\  {\bf 469} (1996) 437 [astro-ph/9603033].

%\cite{Zaldarriaga:1996xe}
\bibitem{Zaldarriaga:1996xe}
M.~Zaldarriaga and U.~Seljak, ``An all sky analysis of polarization in the microwave background,'' Phys.\ Rev.\ D {\bf 55} (1997) 1830 [astro-ph/9609170].

%\cite{Hu:1997hp}
\bibitem{Hu:1997hp}
W.~Hu and M.~J.~White, ``CMB anisotropies: Total angular momentum method,'' Phys.\ Rev.\ D {\bf 56} (1997) 596 [astro-ph/9702170].

%\cite{Weinberg:2008zzc}
\bibitem{Weinberg:2008zzc}
S.~Weinberg, {\it Cosmology\/}, Oxford, UK: Oxford Univ. Pr. (2008) 593 pp.

%\cite{Dodelson:2003ft}
\bibitem{Dodelson:2003ft}
S.~Dodelson, ``Modern Cosmology,'' Amsterdam, Netherlands: Academic Pr. (2003) 440 pp.
  
%\cite{Samaddar}
\bibitem{Samaddar}
S.~N.~Samaddar, ``Some Integrals Involving Associated Legendre Functions,'' Math.\ Comput.\  {\bf 28} (1974) 257.
  
%\cite{Lesgourgues:2011re}
\bibitem{Lesgourgues:2011re}
J.~Lesgourgues, ``The Cosmic Linear Anisotropy Solving System (CLASS) I: Overview,'' arXiv:1104.2932 [astro-ph.IM].

%\cite{Blas:2011rf}
\bibitem{Blas:2011rf}
D.~Blas, J.~Lesgourgues, and T.~Tram, ``The Cosmic Linear Anisotropy Solving System (CLASS) II: Approximation schemes,'' JCAP {\bf 1107} (2011) 034 [arXiv:1104.2933 [astro-ph.CO]].

%\cite{Kanno}
\bibitem{Kanno}
S.~Kanno, ``Cosmological implications of quantum entanglement in the multiverse,'' Phys.\ Lett.\ B {\bf 751} (2015) 316 [1506.07808 [hep-th]]; 
S.~Kanno, ``A note on initial state entanglement in inflationary cosmology,'' Europhys.\ Lett.\  {\bf 111} (2015) no.6, 60007 [1507.04877 [hep-th]].

%\cite{Bolis:2016vas}
\bibitem{Bolis:2016vas}
N.~Bolis, A.~Albrecht and R.~Holman, ``Modifications to Cosmological Power Spectra from Scalar-Tensor Entanglement and their Observational Consequences,'' arXiv:1605.01008 [hep-th].

%\cite{Albrecht:2014aga}
\bibitem{Albrecht:2014aga}
A.~Albrecht, N.~Bolis, and R.~Holman, ``Cosmological Consequences of Initial State Entanglement,'' JHEP {\bf 1411} (2014) 093 [arXiv:1408.6859 [hep-th]].

\end{thebibliography}
\end{document}